\documentclass[twocolumn,superscriptaddress,floats,prd,nofootinbib]{revtex4}
\usepackage{amssymb,amsmath,amssymb,amsfonts,amsthm,stmaryrd,mathrsfs,physics,bbm}
\usepackage{subfigure}
\usepackage{graphicx}
\usepackage[T1]{fontenc}
\usepackage{enumerate} % advanced enumerate environment
\usepackage{color} % for colored text
\usepackage{graphpap} % for numbered coordinate grid
\usepackage{xcolor}
\usepackage[normalem]{ulem}

\usepackage{hyperref} % for HyperTeX cross-referencing

\newcommand{\tx}{\mathfrak{x}}
\newcommand{\sut}{{\rm SU(2)}}
\newcommand{\Meff}{M_{\mathrm{eff}}}

\newcommand{\MB}{M_{\mathrm B}}
\newcommand{\cA}{\mathcal{A}}
\newcommand{\cB}{\mathcal{B}}
\newcommand{\cR}{\mathcal{R}}
\newcommand{\cZ}{\mathcal{Z}}
\newcommand{\BNU}{School of Physics and Astronomy, \mbox{Key Laboratory of Multiscale Spin Physics (Ministry of Education)}, Beijing Normal University, Beijing 100875, China}

\begin{document}

\title{Consistent Gauge Conditions for Dust-Shell Dynamics in Effective Quantum Gravity}

\author{Dongxue Qu}
\affiliation{College of Physics, Chengdu University of Technology, Chengdu, Sichuan 610059, China}

\author{Cong Zhang\footnote{ cong.zhang@bnu.edu.cn}}
\affiliation{\BNU}
\begin{abstract}
Previous analyses of shocks generated by shell-crossing singularities are affected by inappropriate gauge choices, and no systematic method is available for selecting a consistent gauge. To address this issue, we focus on the shock dynamics, which can be effectively described by a thin dust shell interacting with the surrounding dust. As a first step toward the full shell-crossing problem, we neglect this interaction and study an isolated thin shell, for which we develop a systematic method for constructing consistent gauges in generally covariant effective gravity. We apply this method to a generally covariant effective Hamiltonian model of quantum gravity characterized by a quantum parameter $\zeta$, with classical GR recovered in the limit $\zeta\to0$. In this classical limit, the resulting shell dynamics reproduces the Israel junction conditions, providing a nontrivial validation of our method, whereas for $\zeta\neq0$, it exhibits genuine quantum-gravity corrections.
We also show that gauges such as the Painlev\'e-Gullstrand and Schwarzschild ones are incompatible with the presence of a dust shell when imposed on the whole spatial slice. This explains the difficulties in previous treatments. The framework developed here provides a basis for studying shell-crossing singularities and shock dynamics in generally covariant effective black-hole models.
\end{abstract}

%\keywords{Black hole, loop quantum gravity, singularity resolution }

\maketitle
%\tableofcontents

\section{Introduction}\label{sec:introduction}
Black holes (BHs), among the most mysterious objects in the universe, have been extensively studied since they were first predicted \cite{wheeler1968our,PhysRev.56.455}. The singularities at their centers signal the breakdown of classical general relativity (GR) \cite{PhysRevLett.14.57}, and motivate the study of quantum-gravity (QG) effects in black-hole physics \cite{Modesto:2007wp,PhysRevLett.134.101501}. Loop quantum gravity, one of the most promising approaches to quantum gravity \cite{Ashtekar:2004eh,Rovelli:2004tv,Han:2005km,Thiemann:2007pyv}, has been extensively applied in this context \cite{Modesto:2008im,Gambini:2008dy,Corichi:2015xia,BenAchour:2016brs,Ashtekar:2018lag,Zhang:2020qxw,Kelly:2020uwj,Zhang:2021wex,Husain:2021ojz,Alonso-Bardaji:2021yls,Zhang:2021xoa,Munch:2021oqn,Gambini:2022dec,Husain:2022gwp,Alonso-Bardaji:2022ear,Gambini:2022hxr,Lewandowski:2022zce,Alonso-Bardaji:2023vtl,Giesel:2023hys,Giesel:2023tsj,AlonsoBardaji:2023bww,Fazzini:2023ova,Zhang:2024khj,Cafaro:2024vrw,Lin:2024flv,Cipriani:2024nhx,Zhang:2024ney,Lin:2024beb,Yang:2025ufs,Fazzini:2025ysd,Liu:2025fil,Fazzini:2025zrq,Sahlmann:2025fde}. Following many studies of regular black-hole solutions in vacuum, attention has gradually shifted toward their formation through the gravitational collapse of dust \cite{Husain:2021ojz,Husain:2022gwp,Lewandowski:2022zce,Fazzini:2023ova,Giesel:2023hys,Giesel:2023tsj,Cipriani:2024nhx,Shi:2024vki,Fazzini:2025ysd,Liu:2025fil,Fazzini:2025zrq}. Recently, attention has turned to the occurrence of shell-crossing singularities and their dynamics. During the collapse of a dust ball with an inhomogeneous density profile, adjacent dust shells may develop different velocities, so that their worldlines intersect, leading to the formation of a shell-crossing singularity \cite{Fazzini:2023ova,Cipriani:2024nhx,Fazzini:2025ysd,Liu:2025fil,Fazzini:2025zrq,Bobula:2026zlq}. Although the existence of shell-crossing singularities is widely accepted, their dynamics remains an open question \cite{Fazzini:2023ova,Liu:2025fil,Sahlmann:2025fde}.

In classical GR, issues such as shell-crossing singularities are treated using the Israel junction conditions, which provide a weak formulation of the Einstein equations \cite{Israel:1966rt}. In quantum black-hole models, especially those formulated within the Hamiltonian framework, junction conditions have been considered by integrating the effective constraints across the shell \cite{Fazzini:2025nse,Bobula:2026zlq}. However, a systematic generalization of the Israel junction conditions that incorporates the full set of effective field equations, including both the constraints and Hamilton’s equations,  has not yet been established. The lack of such a general framework constitutes the main obstacle to addressing these problems. In previous works, issues related to shell-crossing singularities have typically been addressed by fixing a gauge, that is, by choosing coordinates such as the Painlev\'e-Gullstrand (PG) coordinates, and then solving the Hamilton equations using the Rankine–Hugoniot conditions \cite{Husain:2022gwp,Cipriani:2024nhx}. However, it is often not checked whether the chosen gauge-fixing condition is compatible with the presence of a dust shell-crossing singularity. In the absence of a covariant junction condition analogous to the Israel condition, the shell dynamics must be studied directly from the Hamilton equations together with the Rankine–Hugoniot conditions, which requires a coordinate system that is well defined across the shell. However, it is well known in classical GR that commonly used coordinates, such as the PG and Schwarzschild coordinates, are not continuous at the dust shell \cite{Israel:1966rt,Friedman:1997fu}. 
This shows that such coordinates are not suitable for analyzing the shell dynamics within the Hamiltonian framework. More generally, coordinate artifacts are a common source of confusion in GR. For example, some discontinuity effects predicted in \cite{Husain:2021ojz} were later shown to be purely coordinate artifacts \cite{Fazzini:2023scu}, highlighting the importance of choosing coordinates with care.

The main goal of this work is to develop a framework for choosing appropriate coordinates in the presence of a dust shell. Within the Hamiltonian framework, choosing a coordinate system corresponds to specifying the lapse function and shift vector that determine the dynamics. Our goal is to derive equations that constrain their choice. 
Our analysis shows that imposing gauges such as the PG or Schwarzschild gauges on the entire spatial slice, where the areal radius $g_{\theta\theta}$ is assumed to vary smoothly with the coordinate, is not compatible with the presence of a dust shell, since they lead to ill-defined expressions involving the division by zero.

To derive such equations, a natural starting point is to write down the full set of constraints, including both the gravitational and shell contributions, and then analyze these constraints together with the Hamilton equations, as done, for example, in \cite{Louko:1997wc,Friedman:1997fu}. This approach requires the constraint contribution of the dust shell to be specified explicitly. In the case of interest here, namely when a shell-crossing singularity forms during the collapse of an inhomogeneous dust ball, this becomes more complicated because the resulting dust shell interacts with the surrounding collapsing dust \cite{Sakai:1993vn,Bobula:2026zlq}. This motivates an alternative approach in which the explicit form of the shell contribution to the constraints is not required.

In this work, we follow an idea inspired by classical GR: instead of using the explicit shell contribution to the constraints, we characterize the dust shell through its energy-momentum tensor. This requires identifying the tensor that plays the role of the Einstein tensor in the effective quantum black-hole model within the Hamiltonian framework \cite{DelAguila:2025pgy}. We refer to this tensor as the effective Einstein tensor. The full set of equations can then be combined into a single tensorial equation. By appropriately projecting this equation and using only the assumed properties of the shell energy-momentum tensor, the explicit shell contribution can be eliminated from the resulting junction equations. This allows us to derive the conditions constraining the gauge choice, together with the jump conditions for the regular parts of the canonical variables across the shell, without specifying the explicit form of the shell contribution to the constraints. With these junction conditions determined, the dynamics can then be solved. The remaining components of the effective Einstein tensor can subsequently be evaluated across the shell. Their distributional contributions determine the corresponding components of the shell energy-momentum tensor, thereby providing quantitative information about the dust shell, such as its surface energy density. Conceptually, our strategy is therefore to characterize the matter content of the dust shell through the distributional part of the effective Einstein tensor, rather than by prescribing its canonical contribution to the constraints a priori.

One further point should be emphasized. The above strategy applies only when the theory is covariant, since only then can one define an effective Einstein tensor and rewrite the equations of motion as an Einstein equation. In the Hamiltonian framework, general covariance implies that different choices of lapse function and shift vector lead to solutions that differ only by a spacetime diffeomorphism. The problem of restoring general covariance in the Hamiltonian framework has been systematically addressed in \cite{Zhang:2024khj,Zhang:2024ney,Yang:2025ufs}. In a subsequent work \cite{Zhang:2025ccx}, it was shown that for any static spherically symmetric metric depending on a black-hole mass parameter, one can construct a generally covariant Hamiltonian model in which the given metric is the unique solution. Therefore, the strategy developed in this work applies to a broad class of effective black-hole models.

This paper is organized as follows. In Sec. \ref{sec:pre}, we introduce the phase-space structure of the effective quantum gravity models considered here, together with our minimal assumptions on the energy-momentum tensor of the dust shell. In Sec. \ref{sec:et}, we derive the equations constraining the gauge choice from these assumptions. In Sec. \ref{sec:bardeeneff},  we apply this method to a generally covariant Hamiltonian model of quantum gravity, which admits the Bardeen metric as its unique vacuum solution. Using these equations, we solve the dynamics with numerics in Sec. \ref{sec:dyn}.     Then, the classical Israel junction condition is used as a diagnostic of our numerical results in Sec.~\ref{sec:isseal}. Finally, we present our conclusions and outlook in Sec. \ref{sec:con}.

\section{Preliminary}\label{sec:pre}
The approach developed in this work applies to a broad class of effective quantum gravity models formulated within the Hamiltonian framework. These models have similar phase space structures, which are defined as follows. 

\subsection{Phase-space structure in the vacuum sector}
Let $\Sigma$ be the spatial manifold on which the phase-space variables are defined as fields. Imposing spherical symmetry, we assume that $\Sigma \cong \mathbb X \times \mathbb S^2$, where $\mathbb X$ is a one-dimensional manifold and $\mathbb S^2$ is the two-sphere. The manifold $\Sigma$ is equipped with the $\sut$ action. Let $(x,\theta,\phi)$ denote coordinates on $\Sigma$ adapted to this $\sut$ action. The phase space for gravity consists of fields $(K_I, E^I)$ with $I=1,2$ defined on $ \mathbb X$. Geometrically, $K_I$ is related to the extrinsic curvature of $\Sigma$, while $E^I$ defines the spatial metric on $\Sigma$ (see, e.g., \cite{Bojowald:2004af,Zhang:2021xoa,Gambini:2022hxr} for more details of the kinematical structure). The non-vanishing Poisson brackets between the phase space variables are
\begin{equation}
\begin{aligned}
\{K_1(x),E^1(y)\}=&2\delta(x,y),\\
\{K_2(x),E^2(y)\}=&\delta(x,y),
\end{aligned}
\end{equation}
where the geometrized units $G=1=c$ are used.  

The system is a totally constrained first-class system, whose dynamics is encoded in a set of constraints: the diffeomorphism constraint $H_x$ and the Hamiltonian constraint $H$. For all models considered here, the diffeomorphism constraint takes the same form as in classical GR,
\begin{equation}
H_x=-\frac{1}{2}K_1\partial_xE^1+E^2\partial_xK_2,
\end{equation}
where $H_x$ generates spatial diffeomorphisms. By contrast, the explicit form of the Hamiltonian constraint depends on the model considered (see, e.g., \cite{Zhang:2024khj,Belfaqih:2024vfk}). In this work, we consider only those models for which the constraint algebra closes according to the hypersurface deformation algebra:
\begin{equation}\label{eq:constraintalgebra}
\begin{aligned}
\{H_x[N^x_1],H_x[N^x_2]\}&=H_x[N^x_1\partial_xN^x_2-N^x_2\partial_xN^x_1],\\
\{H[N],H_x[N^x]\}&=-H[N^x\partial_xN],\\
\{H[N_1],H[N_2]\}&=H_x[S(N_1\partial_xN_2-N_2\partial_xN_1)],
\end{aligned}
\end{equation}
where $N$, $N^x$, $N_1$, and $N_2$ are arbitrary smearing functions, and the structure function $S$ is given by $S=E^1/(E^2)^{2}$. We use the abbreviation $F[g]=\int_{\Sigma}F(x)g(x)\,\dd x$.

As a totally constrained system, the dynamics is determined only after specifying a lapse function $N$ and a shift vector $N^x$ to smear the constraints, yielding the Hamiltonian
\begin{equation}\label{eq:Hamiton}
\mathbb H = H[N] + H_x[N^x].
\end{equation}
One may then solve Hamilton's equations generated by $\mathbb H$ to obtain the evolution of the phase-space variables. As proved in \cite{Zhang:2024khj,Zhang:2024ney}, for models whose constraint algebra takes the form given in \eqref{eq:constraintalgebra}, 
the requirement of general covariance uniquely determines the spacetime metric. It takes the form
\begin{equation}\label{eq:metric}
\dd s^2 = -N^2 \dd t^2 + \frac{(E^2)^2}{E^1}\left(\dd x + N^x \dd t\right)^2 + E^1 \dd\Omega^2
\end{equation}
where
$\dd\Omega^2 := \dd\theta^2 + \sin^2\theta\,\dd\phi^2$
denotes the metric on $\mathbb S^2.$ 
As mentioned in Sec.~\ref{sec:introduction}, we need to derive the effective Einstein tensor. Currently, this can only be done for models whose spacetime metric takes the above form. In the following, we work in the triad branch $E^2>0$.

\subsection{Minimal assumptions on dust shell}\label{sec:assumdust}
As discussed in Sec.~\ref{sec:introduction}, we use the energy-momentum tensor to characterize the dust shell. As in the classical theory, the energy-momentum tensor of the dust shell takes the form $T^{\mu\nu} \propto u^\mu u^\nu,$ where $u^\mu$ is the four-velocity of the dust particles. This is our first assumption. In addition, we assume that the dust shell couples only to the 4-dimensional metric \eqref{eq:metric}, rather than to its derivatives, namely the variables $K_I$. This is our second assumption.

Our derivation uses only these two assumptions and does not require the explicit form of the shell contribution to the constraints.

\section{The equation constraining the choice of gauge} \label{sec:et}

As mentioned in Sec.~\ref{sec:introduction}, we need to derive the effective Einstein tensor within the Hamiltonian formulation. To this end, we first reconstruct the action from the constraints, following the procedure of \cite{DelAguila:2025pgy}. 

In the spherically symmetric model, the action reads
\begin{equation}
\begin{aligned}
S=&\int \dd t\,\dd x\left(-\frac{1}{2}K_1\dot E^1-K_2\dot E^2-NH-N^xH_x\right)\\
=&\frac{1}{4\pi}\int \dd^4 x\,\sin\theta\left(-\frac{1}{2}K_1\dot E^1-K_2\dot E^2-NH-N^xH_x\right).
\end{aligned}
\end{equation}
From the action, the effective Einstein tensor is obtained via
\begin{equation}
\delta S=\frac{1}{16\pi}\int \dd^4x\sqrt{-g}G_{\mu\nu}\delta g^{\mu\nu}
\end{equation}
A straightforward calculation gives 
\begin{equation}\label{eq:GinEK}
\begin{aligned}
&G_{\mu\nu}\dd x^\mu\dd x^\nu\\
=&-\frac{2N^2 H}{\sqrt{E^1}E^2}\dd t^2-\frac{4NH_x}{\sqrt{E^1}E^2}(\dd x+N^x\dd t)\dd t\\
&+\frac{2(E^2)^2D_2}{NE^1\sqrt{E^1}}\left(\dd x+N^x\dd t\right)^2+\frac{E^I D_I}{N\sqrt{E^1}E^2}\dd\Omega^2,
\end{aligned}
\end{equation}
where 
\begin{equation}
D_I=-\partial_tK_I+\{K_I, H[N]+H_x[N^x]\},\quad I=1,2,
\end{equation}
and the repeated index $I$ in $E^ID_I$ is summed over $I=1,2$. In \eqref{eq:GinEK}, the effective Einstein tensor is expressed as a function of the 4-dimensional metric, $K_I$, and their derivatives. Once $K_I$ are written in terms of $E^I$ and their derivatives, it can be rewritten as in classical GR so that it depends only on the 4-dimensional metric and its derivatives. To this end, we consider the equations
\begin{equation}\label{eq:derivativesEI}
\partial_t E^I=\{E^I,\mathbb H\},\qquad I=1,2,
\end{equation}
which relate $K_I$ to $E^I$ and their derivatives. Importantly, \eqref{eq:derivativesEI} remains valid even when the dust shell is coupled. This follows from our assumption that the dust shell couples only to the 4-dimensional metric, so that $K_I$ do not appear in the shell part of the constraints. As a result, the shell contribution to the constraints has vanishing Poisson brackets with $E^I$.

Once the effective Einstein tensor is obtained within the Hamiltonian framework, the equations of motion, including the constraints and Hamilton's equations, can be written as
\begin{equation}\label{eq:eninteinequation}
G_{\mu\nu}=8\pi T_{\mu\nu}\propto \delta(x-\tx)u_\mu u_\nu.
\end{equation}
Here we have used the first assumption, and $\tx$ denotes the radial position of the dust shell. The trajectory of the dust shell is described by $\tx(t)$, so that the 4-velocity takes the form
\begin{equation}
u^\mu \partial_\mu \propto \partial_t+\dot{\tx}\partial_x.
\end{equation}
We further introduce the covectors $n_\mu$ and the vector $m^\mu$ by
\begin{equation}
n_\mu \dd x^\mu=\dot{\tx}\dd t-\dd x,
\qquad
m^\mu \partial_\mu=\partial_\theta.
\end{equation}
Then, Eq.~\eqref{eq:eninteinequation} is equivalent to 
\begin{equation}\label{eq:einstein1}
G_{\mu\nu}n^\nu=0,
\quad
G_{\mu\nu}m^\nu=0.
\end{equation}
together with
\begin{equation}\label{eq:einstein2}
G_{\mu\nu} u^\mu u^\nu \propto \delta(x-\tx). 
\end{equation}
We do not consider separately the $\phi$-component of Eq.~\eqref{eq:eninteinequation}, since spherical symmetry ensures that it is proportional to the $\theta$-component.

As in classical GR, suitable continuity conditions must be imposed. For instance, the metric may be required to be continuous across the shell. Once these conditions are specified, the two equations \eqref{eq:einstein1} and \eqref{eq:einstein2} can be solved. 

It should be noted that Eqs. \eqref{eq:einstein1} and \eqref{eq:einstein2} play different roles. Equation \eqref{eq:einstein1} involves only the gravitational degrees of freedom and is independent of the dust shell profile. It therefore determines how the gravitational variables jump across the shell. Once these jumps are determined, they can be substituted into Eq.~\eqref{eq:einstein2}, which relates the dust shell profile to the discontinuity of the gravitational sector.
 
This suggests that our primary focus should be on Eq.~\eqref{eq:einstein1}. By analyzing this equation under the imposed continuity conditions and requiring them to be preserved under evolution, we can derive equations constraining the choice of gauge.

In the following, we illustrate the above analysis using a model of an isolated dust shell coupled to a generally covariant effective Hamiltonian model of quantum gravity. In the vacuum sector, this quantum gravity model admits the Bardeen metric as its unique solution. The analysis also provides a basis for investigating the dynamics of shocks generated by shell-crossing singularities during the collapse of a dust ball. Indeed, such a shock can be regarded as a thin shell interacting with the surrounding dust. The isolated thin-shell model therefore captures the essential shell dynamics while avoiding the additional technical complications associated with this interaction.

\section{Application to a generally covariant effective model}\label{sec:bardeeneff}

In Ref.~\cite{Zhang:2025ccx}, a general procedure was developed to construct an effective Hamiltonian constraint from a prescribed metric, such that the resulting model admits the prescribed metric as its unique vacuum solution. The construction is based on imposing general covariance directly at the level of the $3+1$ Hamiltonian formulation. 

In the current work,  we will consider the Bardeen metric whose line element reads
\begin{equation}
\label{eq:bardeen-metric}
\dd s^2=-f_{\mathrm B}(r;m)\dd t^2
+\frac{\dd r^2}{f_{\mathrm B}(r;m)}
+r^2\dd\Omega^2,
\end{equation}
with
\begin{equation}
f_{\mathrm B}(r;m)
=1-\frac{2mr^2}{(r^2+\zeta^2)^{3/2}},
\end{equation}
where $m$ denotes the mass parameter and $\zeta$, proportional to the Planck length, is a fixed quantum parameter. Applying the reconstruction  procedure in \cite{Zhang:2025ccx} to the Bardeen metric, we can get the effective Hamiltonian constraint $H$ as (see Appendix \ref{app:geteffH} for the detailed derivations)
\begin{equation}
\begin{aligned}
H=&-2E^2\Bigg[
 \cA'(E^1)\left(1-\frac{1}{4}\left(\frac{\partial_xE^1}{E^2}\right)^2\right)\\
 &+\cB'(E^1)(K_2)^2+\cB(E^1) \frac{K_1K_2}{E^2}\\
 &-\frac{\cA(E^1)}{2}\frac{1}{E^2}\partial_x\left(\frac{\partial_xE^1}{E^2}\right)
 \Bigg]
\end{aligned}
\end{equation}
where the functions $\cA$ and $\cB$ are defined as 
\begin{equation}\label{eq:AB-definitions}
\begin{aligned}
 \cA(s)=&\frac{(s+\zeta^2)^{3/2}}{2s},\\
 \cB(s)=&\frac{s^2}{2(s+\zeta^2)^{3/2}}.
\end{aligned}
\end{equation}
This Hamiltonian constraint reduces to that of classical GR in the limit $\zeta\to0$ (see Appendix~\ref{app:GR} for details).

For this Hamiltonian constraint, the evolution equations for $E^I$ given in \eqref{eq:derivativesEI}, which remain valid even when the dust shell is coupled, take the form:
\begin{align}
 \partial_tE^1=&N^x\partial_xE^1+4N\cB K_2,
 \label{eq:E1-evolution}\\
 \partial_tE^2=&\partial_x(N^xE^2)
 +2N\cB K_1+4NE^2\cB'K_2.
 \label{eq:E2-evolution}
\end{align}
Here and throughout the following discussion, unless otherwise specified, $\cA$, $\cB$, and their derivatives are understood to be evaluated at $E^1$.
The Poisson brackets between $K_I$ and $\mathbb H$ read as
\begin{equation}\label{eq:K12-evolution}
\begin{aligned}
 \{K_1,\mathbb H\}=&\partial_x(N^xK_1)
 +2N\Bigg[
 -2E^2\cA''
 +\frac{\cA''(\partial_xE^1)^2}{2E^2}\\
 &-2E^2\cB''(K_2)^2
 -2\cB'K_1K_2
 +\cA'\partial_x(\frac{\partial_xE^1}{E^2})\Bigg]\\
 &+2\partial_xN\left[
 \frac{\cA'\partial_xE^1}{E^2}
 -\frac{\cA\partial_xE^2}{(E^2)^2}
 \right]+\frac{2\cA}{E^2}\partial_x^2N,\\
%%%%%%
\{K_2,\mathbb H\}=&N^x\partial_xK_2
 +N\left[
 -2\cA'-2\cB'(K_2)^2
 +\frac{\cA'(\partial_xE^1)^2}{2(E^2)^2}
 \right]\\
& +\frac{\cA\partial_xE^1\partial_xN}{(E^2)^2}.
\end{aligned}
\end{equation}
Importantly, when the dust shell is coupled, the Poisson brackets are no longer equal to $\partial_t K_I$. 

\subsection{The jump condition}\label{sec:jumpcondition}
Substituting the above results into \eqref{eq:GinEK}, we can now obtain the explicit expression of $G_{\mu\nu}$. As in classical GR, the continuity conditions require the spacetime metric itself to be continuous across the shell, namely
\begin{equation}
 [E^1]=[E^2]=[N]=[N^x]=0,
 \label{eq:metric-continuity}
\end{equation}
while $K_I$ and the first spatial derivatives of $E^I$, $N$, and $N^x$ may
jump.  As a consequence, only second derivatives of the metric can give rise to $\delta$-function terms.
Omitting terms that do not contain $\delta$-function contributions, we
obtain
\begin{equation}
\label{eq:bardeen-Gn}
\begin{aligned}
&\dd x^\mu G_{\mu\nu} n^\nu\\
=&\Bigg(
-\frac{(E^1+\zeta^2)^{3/2}}{(E^1)^{3/2}}
 \frac{\partial_x^2E^1(N^x+\dot\tx)}{(E^2)^2}\\
&-\frac{2N\sqrt{E^1}\partial_xK_2}{(E^2)^2} -\frac{2N^x\left(\dot\tx\,\partial_xK_2+\partial_tK_2\right)}
 {N\sqrt{E^1}}
\Bigg)\dd t\\
&- \frac{2\left(\dot\tx\,\partial_xK_2+\partial_tK_2\right)}
 {N\sqrt{E^1}}\dd x
\\
&+\text{terms without $\delta$-function contributions}.
\end{aligned}
\end{equation}
In the limit $\zeta\to0$, Eq.~\eqref{eq:bardeen-Gn} reduces to the expression for GR.

For a piecewise smooth function $f(t,x)$, the distributional derivatives can be decomposed into regular and singular parts:
\begin{equation}\label{eq:partialxfdeltadt}
\begin{aligned}
\partial_x f
&=(\partial_x f)_{\rm reg}+[f] \delta(x - \tx),\\
\partial_t f
&=(\partial_t f)_{\rm reg}-[f] \dot \tx \delta(x - \tx)
\end{aligned}
\end{equation}
where 
$$
[f]:=f(t,\tx(t)+\epsilon)-f(t,\tx(t)-\epsilon),
$$  
denotes the jump of the quantity $f$ across the shell. Since
we only focus on the delta-function contributions below, the regular parts will
be omitted.

Using these relations, we can simplify \eqref{eq:bardeen-Gn} to obtain 
\begin{equation}
\label{eq:bardeen-jump1}
 \left(\hat N^x+\dot\tx\right)[\partial_xE^1]
 +\frac{2\hat N(\hat E^1)^2}{(\hat E^1+\zeta^2)^{3/2}}[K_2]=0.
\end{equation}
where $\hat g=g(t,\tx(t))$ denotes the value of any quantity $g$ at the location of the dust shell. Following the same procedure,  $G_{\mu\nu}m^\nu=0$ leads to 
\begin{equation}
\label{eq:bardeen-jump-2}
\begin{aligned}
&-\hat E^1\left(\hat N^x+\dot\tx\right)[K_1]
-\hat E^2\left(\hat N^x+\dot\tx\right)[K_2]\\
=&\frac{(\hat E^1+\zeta^2)^{3/2}}{\hat E^2}[\partial_xN]
+\frac{\hat N\sqrt{\hat E^1+\zeta^2}(\hat E^1-2\zeta^2)}
 {2\hat E^1\hat E^2}[\partial_xE^1].
\end{aligned}
\end{equation}
Imposing the continuity condition, the quantities $E^I$, $N$, and $N^x$, as components of the metric, must remain continuous under evolution. Since $N$ and $N^x$ are freely chosen Lagrange multipliers, the nontrivial requirement is therefore that $E^I$ remain continuous throughout the evolution. In other words,  $[E^I]=0$ must remain valid along the moving shell, namely
\begin{equation}
 0=\frac{\dd}{\dd t}[E^I]
 =[\partial_tE^I]+\dot\tx[\partial_xE^I].
 \label{eq:preserve-continuity}
\end{equation}
From the evolution equations \eqref{eq:E1-evolution}--\eqref{eq:E2-evolution},
and using the continuity of $E^I$, $N$, and $N^x$, we have
\begin{align}
[\partial_tE^1]={}&
\hat N^x[\partial_xE^1]
+\frac{2\hat N(\hat E^1)^2}{(\hat E^1+\zeta^2)^{3/2}}[K_2],
\label{eq:E1-time-jump}\\
[\partial_tE^2]={}&
\hat N^x[\partial_xE^2]+\hat E^2[\partial_xN^x]
+\frac{\hat N(\hat E^1)^2}{(\hat E^1+\zeta^2)^{3/2}}[K_1]\notag\\
&+\frac{\hat N\hat E^1\hat E^2(\hat E^1+4\zeta^2)}
 {(\hat E^1+\zeta^2)^{5/2}}[K_2].
\label{eq:E2-time-jump}
\end{align}
Substituting Eq.~\eqref{eq:E1-time-jump} into
Eq.~\eqref{eq:preserve-continuity} gives Eq.~\eqref{eq:bardeen-jump1}.
Substituting Eq.~\eqref{eq:E2-time-jump} gives
\begin{equation}
\label{eq:bardeen-jump-3}
\begin{aligned}
0={}&\left(\hat N^x+\dot\tx\right)[\partial_xE^2]
+\frac{\hat N(\hat E^1)^2}{(\hat E^1+\zeta^2)^{3/2}}[K_1]\\
& +\frac{\hat N\hat E^1\hat E^2(\hat E^1+4\zeta^2)}
 {(\hat E^1+\zeta^2)^{5/2}}[K_2]
+\hat E^2[\partial_xN^x].
\end{aligned}
\end{equation}
Solving Eqs.~\eqref{eq:bardeen-jump1}, \eqref{eq:bardeen-jump-2} and \eqref{eq:bardeen-jump-3} gives
\begin{align}
 \hat N^x+\dot\tx=&
 -\frac{2\hat N(\hat E^1)^2[K_2]}
 {(\hat E^1+\zeta^2)^{3/2}[\partial_xE^1]},
 \label{eq:bardeen-jump-solved-v}\\
 [\partial_xN^x]=&
 \frac{2\hat N(\hat E^1)^2[\partial_xE^2][K_2]}
 {\hat E^2(\hat E^1+\zeta^2)^{3/2}[\partial_xE^1]}-\frac{\hat N(\hat E^1)^2}
 {\hat E^2(\hat E^1+\zeta^2)^{3/2}}[K_1]\notag\\
 &-\frac{\hat N\hat E^1(\hat E^1+4\zeta^2)}
 {(\hat E^1+\zeta^2)^{5/2}}[K_2],
 \label{eq:bardeen-jump-solved-shift}\\
 [\partial_xN]=&
 \frac{2\hat N\hat E^2(\hat E^1)^2[K_2]}
 {(\hat E^1+\zeta^2)^3[\partial_xE^1]}
 \left(\hat E^1[K_1]+\hat E^2[K_2]\right)\notag\\
 &- \hat N
 \frac{\hat E^1-2\zeta^2}{2\hat E^1(\hat E^1+\zeta^2)}
 [\partial_xE^1].
 \label{eq:bardeen-jump-solved-lapse}
\end{align}

\section{Solving the dynamics}\label{sec:dyn}
To solve the dynamics, it is convenient to choose a gauge, i.e., to fix the coordinates, such that  $\dot\tx=0$, namely $\tx(t)=\tx_0$ for some fixed $\tx_0$. With this gauge choice, the dynamics can be obtained by solving the vacuum equations of motion separately in the two regions $x<\tx_0$ and $x>\tx_0$:
 \begin{equation}\label{eq:hamiltoneqvacuum}
\begin{aligned}
\dot E^I&=\{E^I,\mathbb H\},\\
\dot K_I&=\{K_I,\mathbb H\}. 
\end{aligned}
\end{equation}
Here the lapse function $N$ and the shift vector $N^x$ must be chosen such that the jump conditions \eqref{eq:bardeen-jump-solved-v}--\eqref{eq:bardeen-jump-solved-lapse} are satisfied. We now investigate the jump conditions with $\dot\tx=0$.

With $\dot \tx=0$, Eqs. \eqref{eq:bardeen-jump-solved-v}--\eqref{eq:bardeen-jump-solved-lapse} becomes
\begin{equation}
\label{eq:bardeen-jumpfinal0}
\begin{aligned}
\hat N^x=&
-\frac{2(\hat E^1)^2[K_2]}
 {(\hat E^1+\zeta^2)^{3/2}[\partial_xE^1]},
\\
[\partial_xN^x]=&
-\frac{\hat N^x}{\hat E^2}[\partial_xE^2]
-\frac{(\hat E^1)^2}
 {\hat E^2(\hat E^1+\zeta^2)^{3/2}}[K_1]\\
&-\frac{\hat E^1(\hat E^1+4\zeta^2)}
 {(\hat E^1+\zeta^2)^{5/2}}[K_2],
\\
[\partial_xN]=&
-\frac{\hat E^1\hat E^2\hat N^x}
 {(\hat E^1+\zeta^2)^{3/2}}[K_1]
-\frac{(\hat E^2)^2\hat N^x}
 {(\hat E^1+\zeta^2)^{3/2}}[K_2]\\
 &+\frac{(2\zeta^2-\hat E^1)}
 {2\hat E^1(\hat E^1+\zeta^2)}[\partial_xE^1].
\end{aligned}
\end{equation}
where we set $\hat N=1$ to fix the freedom in the choice of time coordinate. Other choices of $\hat N$, which may depend on $t$, correspond to a rescaling of the time coordinate $t$. Moreover, the last two equations are obtained by 
using the first equation to replace the factors
$[K_2]/[\partial_xE^1]$ in Eqs.~\eqref{eq:bardeen-jump-solved-shift} and
\eqref{eq:bardeen-jump-solved-lapse} by $\hat N^x$. 
We observe that these equations can be satisfied by choosing the lapse function $N$ and the shift vector $N^x$ to obey the following relations in a neighborhood of the dust shell:
\begin{equation}\label{eq:jumpfinal01pbardeen}
\begin{aligned}
\partial_xN^x=&-\frac{N^x}{E^2}\partial_xE^2
-\frac{ (E^1)^2K_1}
 {E^2(E^1+\zeta^2)^{3/2}}\\
&-\frac{ E^1( E^1+4\zeta^2)}
 {( E^1+\zeta^2)^{5/2}}K_2,
\\
\partial_xN=&
N\Bigg(-\frac{N^xE^2 }
 {( E^1+\zeta^2)^{3/2}}\left(E^1K_1+E^2K_2\right)\\
 &+\frac{2\zeta^2-E^1}
 {2 E^1(E^1+\zeta^2)}\partial_xE^1\Bigg).
\end{aligned}
\end{equation}
with boundary conditions 
\begin{equation}\label{eq:junctionCon}
\hat N^x=-\frac{2(\hat E^1)^2[K_2]}
 {(\hat E^1+\zeta^2)^{3/2}[\partial_xE^1]},\quad \hat N=1.
\end{equation}
Here, we include an overall factor of $N$ on the RHS of $\partial_xN$ to ensure that $N$ is positive definite.

In principle, the gauge condition \eqref{eq:jumpfinal01pbardeen} could be imposed on the whole spatial slice $\Sigma$ to solve the dynamics. In practice, however, when solving the equations of motion \eqref{eq:hamiltoneqvacuum} numerically, one must also specify boundary conditions. For this reason, it is preferable to impose the gauge condition \eqref{eq:jumpfinal01pbardeen} only in a neighborhood of the dust shell $x=\tx_0$, while requiring that the gauge reduce to a more familiar one away from the shell. We therefore choose the gauge 
\begin{equation}\label{eq:defineNs6}
\begin{aligned}
N=&g_s\mathring N+\frac{g_b}{2}\frac{\partial_xE^1}{E^2},\\
N^x=&g_s\mathring N^x
 -g_b\frac{(E^1)^2K_2}{E^2(E^1+\zeta^2)^{3/2}}.
\end{aligned}
\end{equation}
Here $\mathring N$ and $\mathring N^x$ are solutions of \eqref{eq:jumpfinal01pbardeen},
 i.e.,
\begin{equation}
\label{eq:bardeen-local-gauge}
\begin{aligned}
 \partial_x\mathring N^x=&
 -\mathring N^x\frac{\partial_xE^2}{E^2}
 -\frac{(E^1)^2}{E^2(E^1+\zeta^2)^{3/2}}K_1\\
& -\frac{ E^1(E^1+4\zeta^2)}
 {(E^1+\zeta^2)^{5/2}}K_2,
\\
 \partial_x\mathring N=&
 \mathring N\Bigg[
 -\frac{\mathring N^xE^2}{(E^1+\zeta^2)^{3/2}}
 \left(E^1K_1+E^2K_2\right)\\
& +\frac{2\zeta^2-E^1}{2E^1(E^1+\zeta^2)}\partial_xE^1
 \Bigg],
\end{aligned}
\end{equation}
The function $g_s$ is chosen such that $g_s(\tx_0)=1$ and $g_s'(\tx_0)=0$, and $g_s(x)=0$ for $|x-\tx_0|\gg 1$, with $g_b=1-g_s.$ This choice ensures that, far away from the dust shell, the gauge reduces to one for which $\dot E^I=0=\dot K_I$.

\subsection{Rewriting the equations of motion for numerical investigation}\label{sec:newvar4num}
To solve the equations of motion numerically, we introduce the variables 
\begin{equation}
\begin{aligned}
&s_1=E^1,\ s_2=K_2,\ s_3=\frac{K_1}{E^2},\ s_4=\frac{\partial_xs_1}{E^2},\\
&s_5=\frac{\partial_xs_4}{E^2},\ s_6=N^x E^2,\ s_7=\frac{\partial_xN}{E^2}\sqrt{s_1}.
\end{aligned}
\end{equation}
We then define $\vec u$ as 
\begin{equation}
\vec u=\left(E^2,E^2s_1,E^2s_4,E^2s_2,E^2s_3\right).
\end{equation}
In terms of the $s$-variables, the equations of motion take the form of a balance law:
\begin{equation}\label{eq:conservation}
\partial_t \vec{u} =\partial_x F(\vec{u}) + J(\vec{u})
\end{equation}
\begin{equation}
\label{eq:bardeen-flux}
F=
\begin{pmatrix}
s_6\\
\\
s_1s_6\\
\\
s_4s_6+4N\cB s_2\\
\\
s_6s_2+N\cA s_4\\
\\
s_6s_3+\dfrac{2\cA}{\sqrt{s_1}}s_7
\end{pmatrix},
\end{equation}
and
\begin{equation}
\label{eq:bardeen-source}
J=
\begin{pmatrix}
NE^2\left(2\cB s_3+4\cB' s_2\right)\\
\\
NE^2\left[2s_1\cB s_3+\left(4s_1\cB'+4\cB\right)s_2\right]\\
\\
0\\
\\
-4NE^2\cA'\\
\\
NE^2\,\mathcal J_{\mathrm B}
\end{pmatrix}.
\end{equation}
Here $\cA,\cB$ and their derivatives are evaluated at $s_1$, and
\begin{equation}
\label{eq:bardeen-JB}
\begin{aligned}
\mathcal J_{\mathrm B}
=&2\bigg[
-2\cA''
+\frac{\cA''(s_4)^2}{2}
-2\cB''(s_2)^2
-2\cB's_2s_3 \\
&+ \frac{2\cA'}{\cA}
\left(
\cA'\left(1-\frac{(s_4)^2}{4}\right)
+\cB'(s_2)^2+\cB s_2s_3
\right)
\bigg].
\end{aligned}
\end{equation}
In addition, the constraints $H=0$ and $H_x=0$ have been used in deriving the above equations. This is valid since we only need to solve the equations of motion in the vacuum regions $x\neq \tx_0$, where the constraints are satisfied.

In terms of the $s$-variables, we can rewrite the lapse function and the shift vector as 
\begin{equation}
\label{eq:bardeen-s-gauge}
N=g_s\mathring N+\frac{g_b}{2}s_4,\  
s_6=g_s\mathring s_6
-g_b\frac{(s_1)^2}{(s_1+\zeta^2)^{3/2}}s_2,
\end{equation}
where we introduce 
\begin{equation}
\mathring s_6=E^2\mathring N^x.
\end{equation}
According to \eqref{eq:bardeen-local-gauge}, $\mathring N$ and $\mathring s_6$ satisfy  
\begin{equation}
\label{eq:bardeen-s-local-gauge}
\begin{aligned}
\frac{\partial_x\mathring N}{E^2}
=&-\mathring N\Bigg[\frac{s_1-2\zeta^2}{2s_1(s_1+\zeta^2)}s_4\\
&
+\frac{\mathring s_6}{(s_1+\zeta^2)^{3/2}}
\left(s_1s_3+s_2\right)
\Bigg],\\
\frac{\partial_x\mathring s_6}{E^2}
=&-\left[
\frac{(s_1)^2}{(s_1+\zeta^2)^{3/2}}s_3
+\frac{s_1(s_1+4\zeta^2)}{(s_1+\zeta^2)^{5/2}}s_2
\right].
\end{aligned}
\end{equation}
with boundary conditions at the shell
\begin{equation}
\label{eq:bardeen-s-gauge-boundary}
\widehat{\mathring s_6}
=-\frac{2(\hat s_1)^2[s_2]}
{(\hat s_1+\zeta^2)^{3/2}[s_4]},
\qquad
\widehat{\mathring N}=1.
\end{equation}
For the scalar $s_7$, we have
\begin{equation}
\label{eq:bardeen-s7}
\begin{aligned}
s_7=&
\sqrt{s_1}\frac{\partial_xg_s}{E^2}
\left(\mathring N-\frac{s_4}{2}\right)\\
&-g_s\mathring N\Bigg[\frac{(s_1-2\zeta^2)s_4}{2\sqrt{s_1}(s_1+\zeta^2)}\\
&+\frac{\sqrt{s_1}\mathring s_6}{(s_1+\zeta^2)^{3/2}}
\left(s_1s_3+s_2\right)
\Bigg]\\
&+g_b\frac{\sqrt{s_1}}{\cA}
\left[
\cA'\left(1-\frac{(s_4)^2}{4}\right)
+\cB'(s_2)^2+\cB s_2s_3
\right].
\end{aligned}
\end{equation}
where we have used $H=0$, which implies
\begin{equation}\label{eq:s5Heffp}
s_5=\frac{2}{\cA}
\left[
\cA'\left(1-\frac{(s_4)^2}{4}\right)
+\cB'(s_2)^2+\cB s_2s_3
\right].
\end{equation}
We now obtain the full set of equations of motion in terms of the $s$-variables, from which we can solve the dynamics. 

\subsection{Numerical results}

Given initial data $K_I$ and $E^I$, and thus $s_a$ with $a=1,2,\cdots,5$, on the initial time slice $t_0$ satisfying the constraints, the first step is to determine $\mathring{s}_6$ and $\mathring{N}$ by solving \eqref{eq:bardeen-s-local-gauge} with the initial condition \eqref{eq:bardeen-s-gauge-boundary}. One can then obtain $N$ and $s_6$ from \eqref{eq:bardeen-s-gauge}, which encode the lapse function and the shift vector. With $N$ and $s_6$ determined, the balance-law system \eqref{eq:conservation} is solved to compute the values of $s_a$ with $a=1,2,\cdots,5$  on the subsequent time slice $t_0+\delta t$. Repeating this procedure yields the data on all subsequent time slices. The system~\eqref{eq:conservation} is solved numerically using a Kurganov--Tadmor (KT) finite-volume scheme with a Runge--Kutta (RK) time integrator and the minmod limiter described in \cite{Liu:2025fil}. The codes used in the current work are available in \cite{github}.

\subsubsection{Initial condition}\label{sec:initial}
We denote by $m_i$ and $m_e$ the mass parameter of the interior and the exterior regions. We choose the initial data as follows:
\begin{itemize}
\item[(1)] for the interior region, we choose
\begin{equation}
(s_1)^{\rm in}=x^2,\quad (E^2)^{\rm in}=x;
\end{equation}
\item[(2)] for the exterior region, we choose 
\begin{equation}\label{eq:ini1}
\begin{aligned}
(s_1)^{\rm ex}=&\big(x+f((x-\tx_0)/l)\big)^2,\\
E^2=&x+f((x-\tx_0)/l),
\end{aligned}
\end{equation}
where $l$ is a small parameter, and
\begin{equation}\label{eq:ini2}
f(x)=a (\tanh(x) + \frac{1}{2}\tanh(x)^2),
\end{equation}
for some constant $a$. 
\end{itemize}

With the chosen expressions for $s_1$ and $E^2$, the value of $s_4$ is obtained by its definition, 
\begin{equation}
s_4=\frac{\partial_xs_1}{E^2}. 
\end{equation}
The remaining variables can then be obtained from the constraints, which yield
\begin{equation}
\MB
=\cA(s_1)\left(1-\frac{(s_4)^2}{4}\right)
+\cB(s_1)(s_2)^2=m_o,
\end{equation}
for some constant $m_o=m_i$ or $m_e$. Here, $m_i$ and $m_e$ are precisely the values of the mass parameter appearing in the Bardeen metric in the interior and exterior regions, respectively. This result leads to
\begin{equation}
\label{eq:bardeen-initial-s2}
s_2=\pm
\sqrt{
\frac{m_o-\cA\left(1-\frac{(s_4)^2}{4}\right)}{\cB}
}.
\end{equation}
The choice of sign distinguishes between the expanding and collapsing branches of the dust-shell dynamics.
Moreover, from the diffeomorphism constraint, we have
\begin{equation}
s_3=\frac{2\partial_xs_2}{E^2 s_4}.
\end{equation}

According to \eqref{eq:bardeen-s-gauge-boundary}, we obtain
\begin{equation}\label{eq:nonareal}
\hat s_6=-\frac{2(\hat s_1)^2[s_2]}
{(\hat s_1+\zeta^2)^{3/2}[s_4]}.
\end{equation}
This result has an important consequence. If the gauge is chosen such that $\partial_x E^1$ is continuous, i.e., $[s_4]=0$, then according to \eqref{eq:bardeen-initial-s2}, we have $[s_2]\neq 0$ whenever the interior and exterior masses are different. Consequently, $\hat s_6$ is not well defined. In other words, for $\hat s_6$ to be well defined, $[s_4]$ must be nonzero; equivalently, $\partial_x E^1$ cannot be continuous across the shell. This explains why commonly used gauges, such as the PG gauge or the Schwarzschild gauge, in which $E^1=x^2$, cannot be used once a dust shell is coupled to the system.
\subsubsection{The results}
\begin{figure*}[!t]
\centering
\begin{minipage}{0.48\textwidth}
    \centering
    \includegraphics[width=\linewidth]{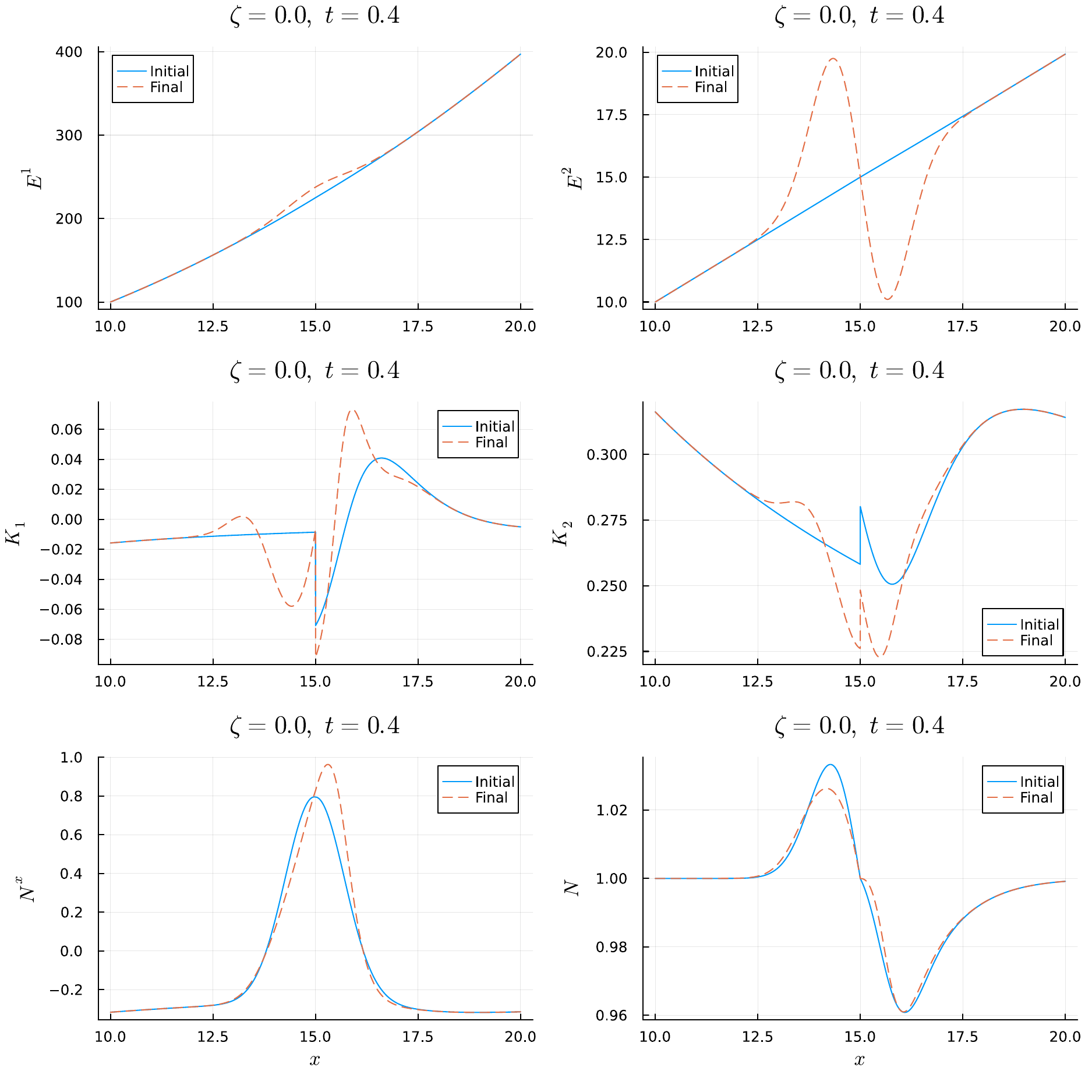}\\
    (a) $\zeta=0$
\end{minipage}
\hfill
\begin{minipage}{0.48\textwidth}
    \centering
    \includegraphics[width=\linewidth]{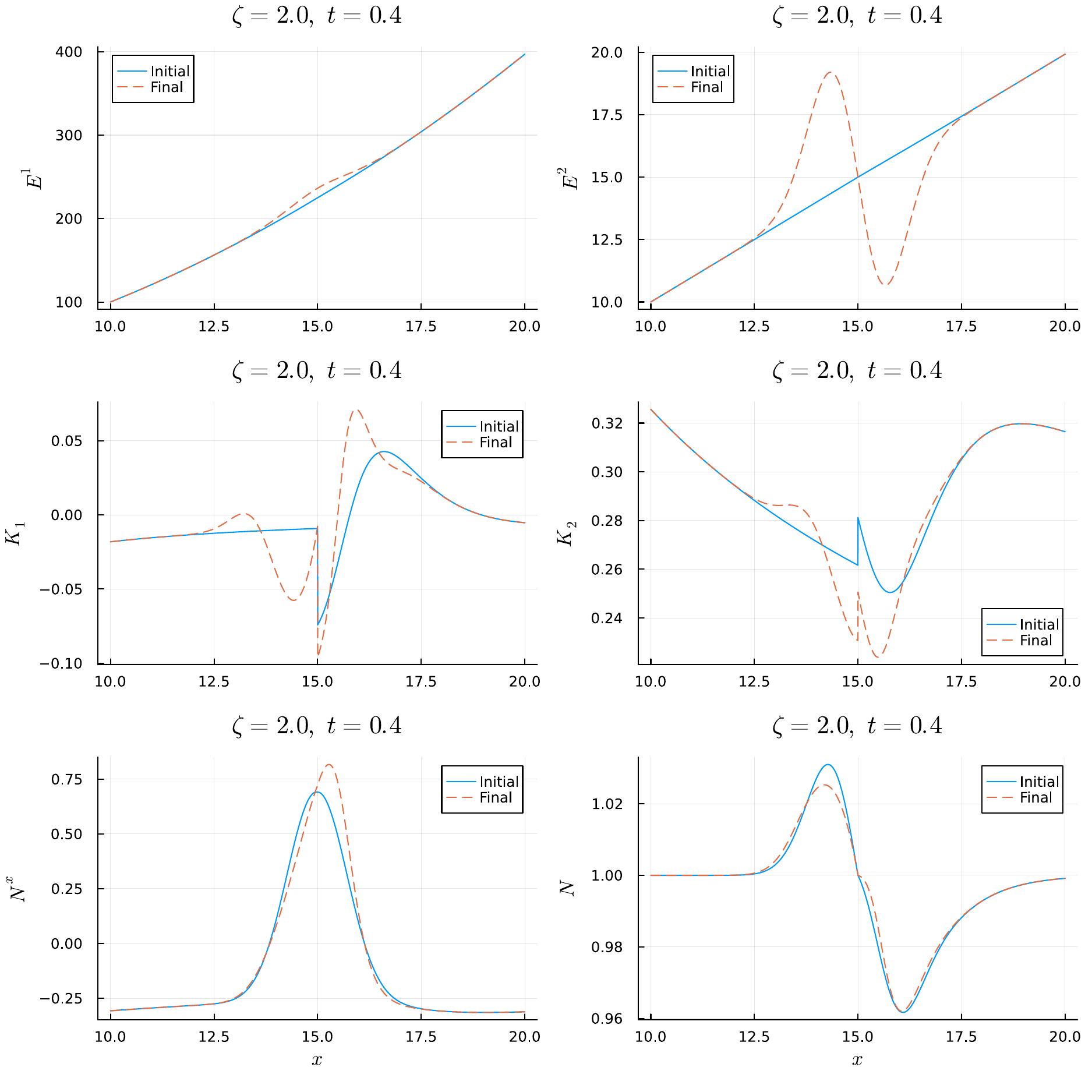}\\
    (b) $\zeta=2$
\end{minipage}
\par\vspace{8mm}
\begin{minipage}{0.48\textwidth}
    \centering
    \includegraphics[width=\linewidth]{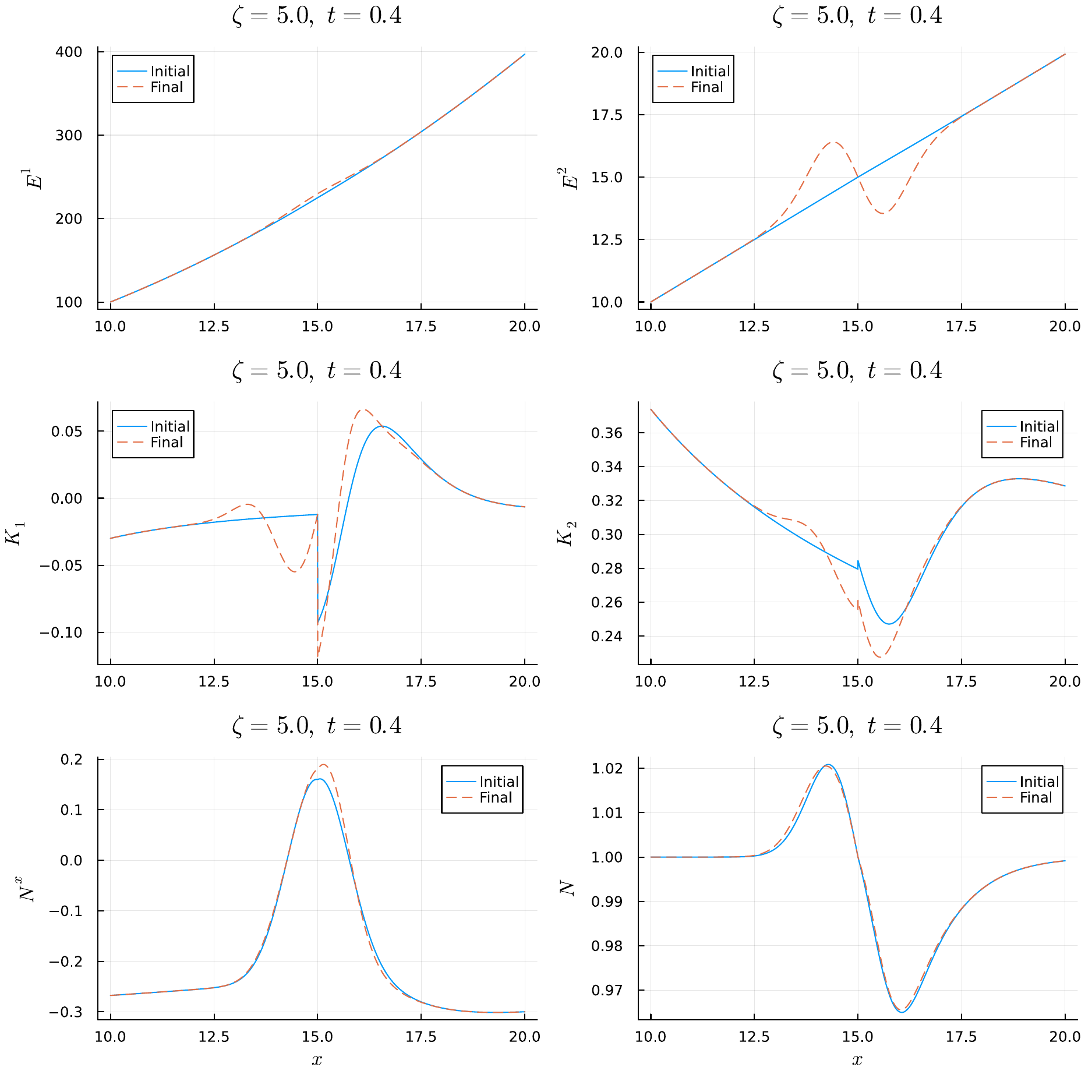}\\
    (c) $\zeta=5$
\end{minipage}
\hfill
\begin{minipage}{0.48\textwidth}
    \centering
    \includegraphics[width=\linewidth]{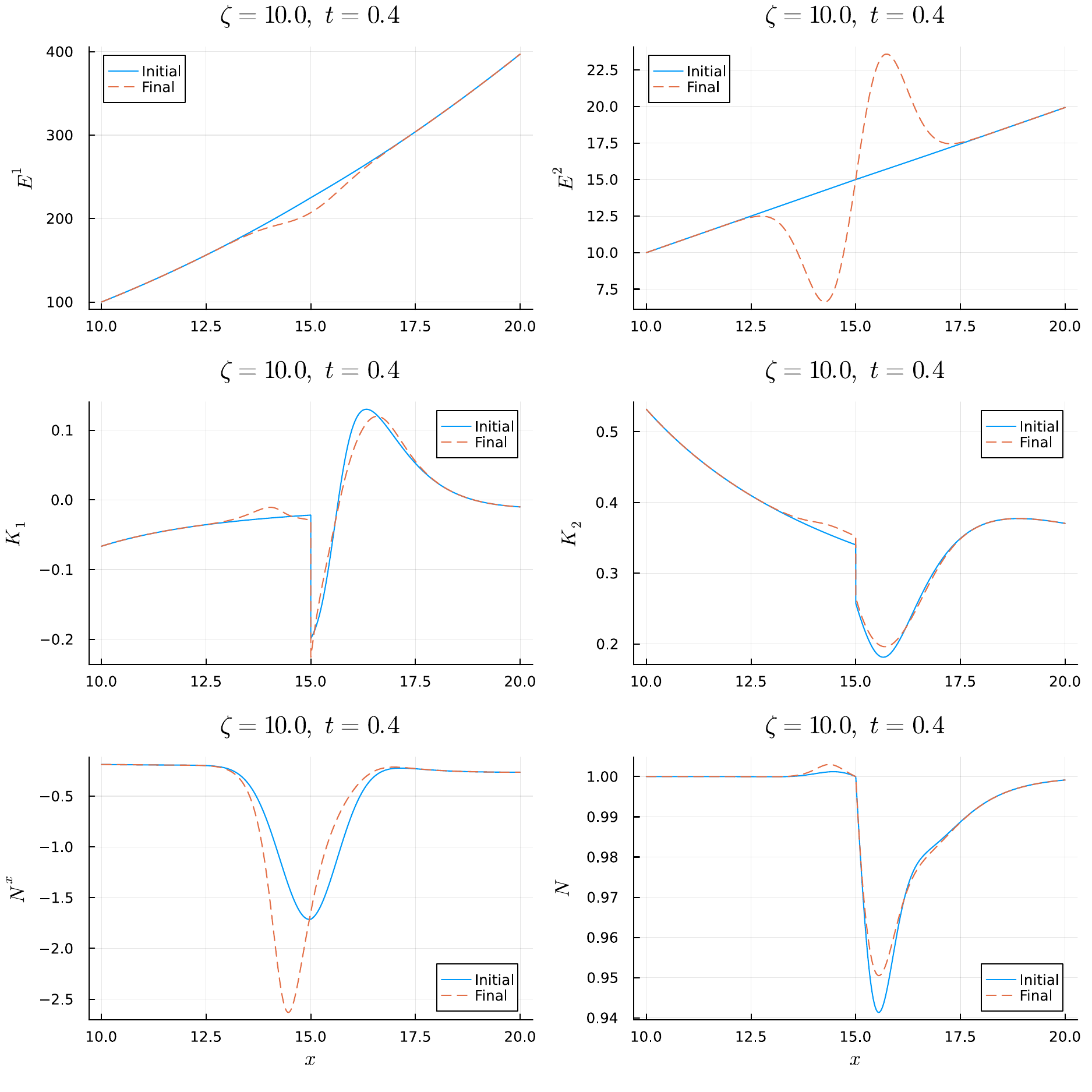}\\
    (d) $\zeta=10$
\end{minipage}
\caption{Initial (solid blue) and evolved (dashed red) profiles of $E^I$, $K_I$, $N$, and $N^x$ for $m_i=1/2$ and $m_e=1$. The cases $\zeta=0$, $2$, $5$ and $10$ are considered.}
\label{fig:num1}
\end{figure*}

Numerical results for $E^I$, $K_I$,  $N$ and $N^x$ are shown in Fig. \ref{fig:num1}, where $\zeta=0$, $2$, $5$ and $10$ are considered.  The result for $\zeta=0$ coincides with that obtained in classical GR (see Appendix \ref{app:GR} and the numerical results therein), while the results for $\zeta\neq 0$ show how quantum-gravity effects modify the dynamics of the dust shell.  As expected, the deviation from the classical GR dynamics becomes increasingly pronounced as $\zeta$ increases.

In the examples, we choose $g_s(x)=\exp(-(x-\tx_o)^2/\sigma^2)$ with $\sigma=1$. The radial position of the dust shell is chosen to be $\tx_o=15$. The parameters $a$ and $l$ introduced in \eqref{eq:ini1} and \eqref{eq:ini2} must be chosen so that the 4-velocity of the dust shell,
$u^\mu \partial_\mu \propto \partial_t+\dot{\hat x}\,\partial_x=\partial_t,$
is timelike. This condition is satisfied for $l=1.8$ and $a=-0.05$. The numerical results indicate that the corresponding initial data describe an expanding, rather than collapsing, dust shell.

\subsection{New Approach Without Separating the Interior and Exterior Regions} 
In the approach described above, we solve the equations \eqref{eq:hamiltoneqvacuum}, which are valid only in the vacuum regions. In other words, these equations are not valid at the location of the dust shell, since the energy-momentum tensor of the shell is not included in the equations of motion. As a result, one must divide the spacetime into two regions, treat the shell as a boundary, and solve the equations separately in the interior and exterior regions. We now introduce an alternative approach in which the dust shell does not need to be treated as a boundary.

Once the dust shell is considered, the equations of motion we should solve become 
\begin{equation}\label{eq:EOMswithmatter}
\begin{aligned}
0&=H+\delta\text{-function term},\\
0&=H_x+\delta\text{-function term},\\
\dot E^I&=\{E^I,\mathbb H\},\\
\dot K_I&=\{K_I,\mathbb H\}+\delta\text{-function term}.
\end{aligned}
\end{equation}
We emphasize that the equations for $\dot E^I$ do not contain $\delta$-function contributions. Indeed, under our assumption that the shell part of the constraints depends only on the metric and the dust-shell degrees of freedom, one has
\begin{equation}
\begin{gathered}
\{E^I,H[N]+H_x[N^x]+\delta\text{-function terms}\}\\
=
\{E^I,H[N]+H_x[N^x]\}.
\end{gathered}
\end{equation}

The main difficulty we now face arises from the $\delta$-function terms appearing in the system. In general, their explicit form is not known, as discussed in Sec.~\ref{sec:introduction}. This obstacle can be overcome by considering suitable linear combinations of the equations such that the $\delta$-function terms cancel. The question is how to identify such combinations without knowing the explicit form of these terms. 

To address this issue, we use the equivalence between the above system and the effective Einstein equation $G_{\mu\nu}=8\pi T_{\mu\nu}$. According to our assumption on $T_{\mu\nu}$, we have
$
\widehat G^{\mu\nu}n_\nu=0,$
for any covector $n_\mu$ satisfying $n_\mu u^\mu=0$, where $\widehat G_{\mu\nu}$ denotes the value of $G_{\mu\nu}$ at the shell. Therefore, the relations
$
\widehat G^{\mu\nu}n_\nu=\widehat T^{\mu\nu}n_\nu=0
$
for different choices of the covector $n_\mu$ provide a set of linear combinations of \eqref{eq:EOMswithmatter}. Since the right-hand side vanishes, these combinations are expected to eliminate the $\delta$-function terms appearing in the equations of motion. 

There is, however, a further issue. The covector $n_\mu$ is defined only along the trajectory of the dust shell, while for our purpose it must be extended to a covector field on the entire spacetime. Under the gauge choice adopted here, one has
$n=\dd x|_{\text{dust shell}}$, so a natural extension is
$\bar n=\dd x$. The condition $G^{\mu\nu}\bar n_\nu=0$ then leads to
\begin{equation}\label{eq:dotK2new0}
\begin{aligned}
\partial_tK_2=&\frac{\partial_xE^1 \partial_xE^2 N \cA}{(E^2)^3}+\frac{\partial_xE^1 \partial_xN \cA}{(E^2)^2}-\frac{\partial_x^2E^1 N \cA}{(E^2)^2}+\\
&\frac{2 K_1K_2 N \cB}{E^2}+\frac{E^1 \partial_xE^1 K_1 N^2}{2 (E^2)^3 N^x}+\\
&\frac{\partial_xE^1 K_1 N^x}{2 E^2}-\frac{E^1 \partial_xK_2 N^2}{(E^2)^2 N^x}
,\\
\partial_tK_2=&\frac{(\partial_xE^1)^2 N \cA'}{2 (E^2)^2}-2 N \cA'+\frac{\partial_xE^1 \partial_xN \cA}{(E^2)^2}-\\
&2K_2^2 N \cB'+\frac{\partial_xE^1 K_1 N^x}{2 E^2}.
\end{aligned}
\end{equation}
These two expressions arise because $G^{\mu\nu}\bar n_\nu$ is non-trivial for $\mu=t,x$. 
Taking the difference of these two expressions, we obtain
\begin{equation}\label{eq:combineconstraint}
\frac{E^1N^2}{(E^2)^3N^x}H_x+\frac{HN}{E^2}=0.
\end{equation}
This result shows explicitly how the Hamiltonian and diffeomorphism constraints can be combined so that the $\delta$-function terms cancel.

For the numerical implementation, one may adopt the second equation in \eqref{eq:dotK2new0} as the evolution equation for $K_2$. Likewise, the relation
$G^{\mu\nu}(\dd\theta)_\mu=0$
provides the corresponding evolution equation for $K_1$. Since $G^{\mu\nu}(\dd\theta)_\mu$ is nontrivial only for $\mu=\theta$, the equation $G^{\mu\nu}(\dd\theta)_\mu=0$ yields a single nontrivial relation. Together with
$\partial_t E^I=\{E^I,\mathbb H\},$
these equations form a complete system of evolution equations that determine the dynamics. 

The above formulation applies directly to the Bardeen effective model considered in this work. As a numerical test, however, we apply it only to the classical GR case, i.e., the case with $\zeta=0$, as shown in Fig.\ref{fig:newapp} (see Appendix\ref{app:GR} for the corresponding equations in GR). Comparing the result with that obtained using the first approach in the same case (see Fig. \ref{fig:num1cl}), we find that the two approaches are in complete agreement.

\begin{figure}[!t]
\centering 
\includegraphics[width=0.45\textwidth]{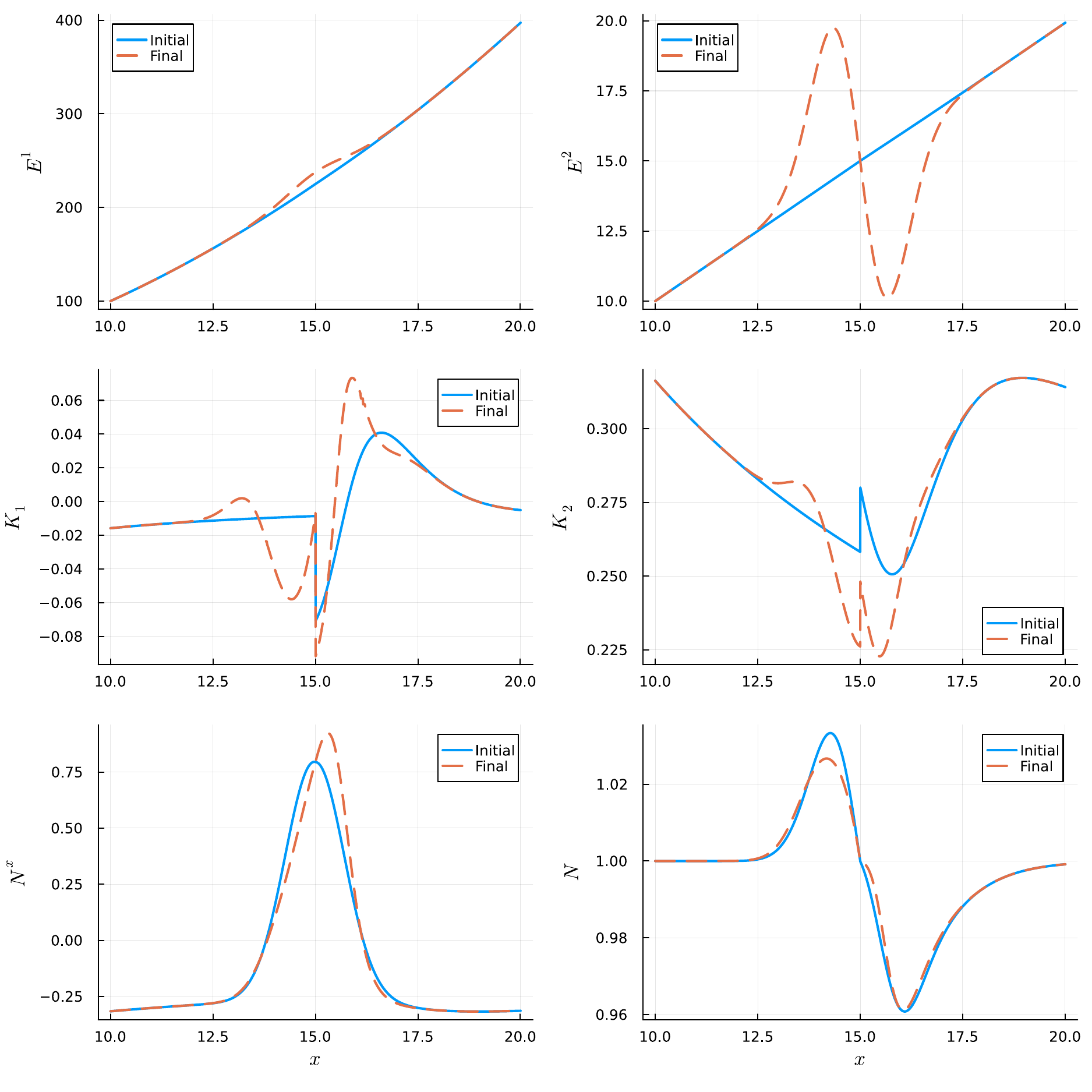}
\caption{Numerical results obtained using the new approach, with the same parameters and initial data as in Fig. \ref{fig:num1cl}, shown at $t=0.4$. The results agree with those obtained from the previous approach up to numerical errors.}\label{fig:newapp}
\end{figure}

\section{Classical Israel junction condition as a diagnostic}\label{sec:isseal}
To validate our numerical strategy, we compare it with the result obtained from the Israel junction condition. In classical GR coupled to a dust shell, the Israel junction condition yields \cite{Israel:1966rt}
\begin{equation}
\begin{aligned}[]
[\partial_xE^1]= \frac{-2\beta \hat E^2   \hat N m_d}{\sqrt{\hat E^1}},
\end{aligned}
\end{equation}
where $m_d$ is the mass of the dust shell and $\beta$ is given by 

\begin{figure*}[t]
\centering
\includegraphics[width=0.8\textwidth]{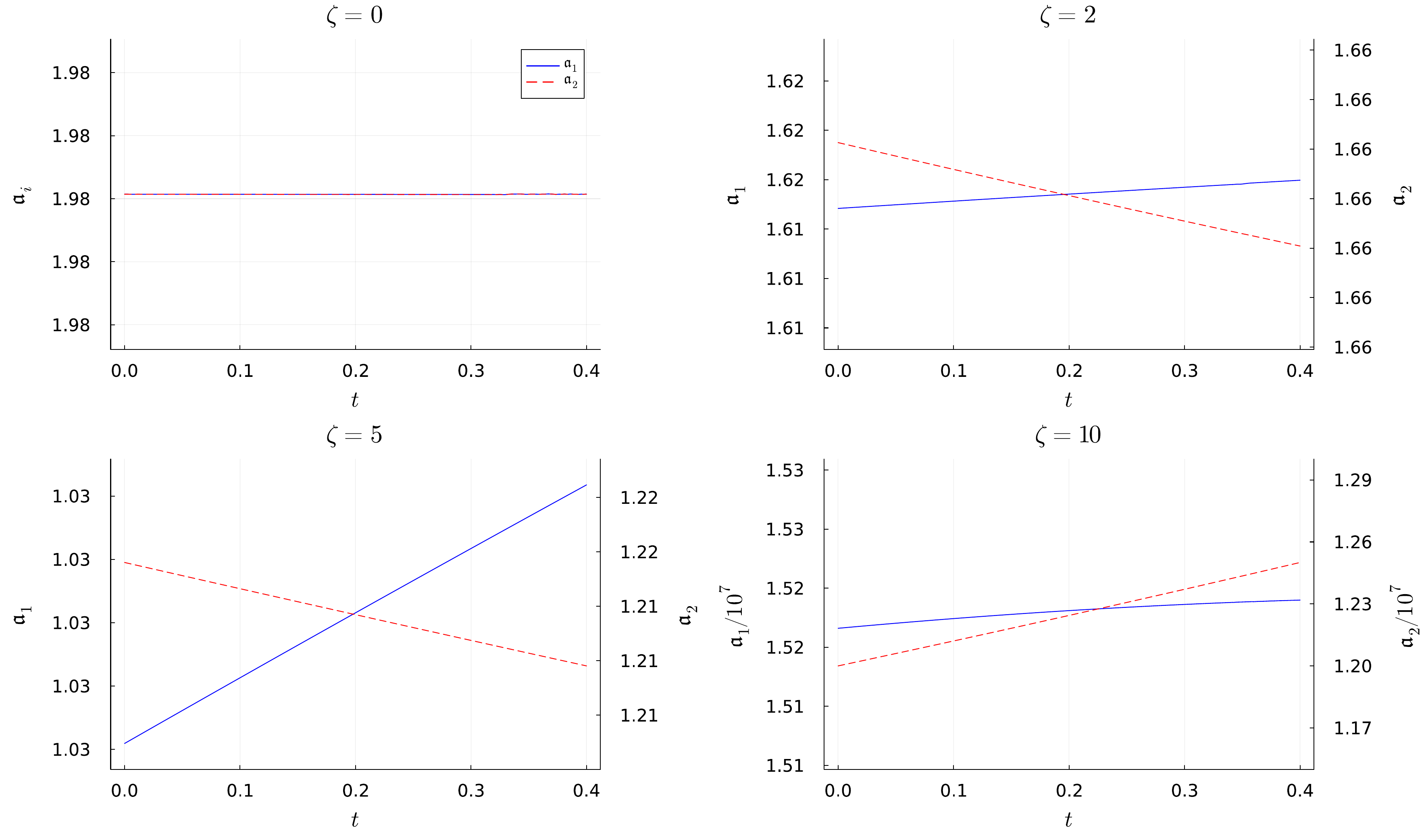}
\caption{Numerical results of $\mathfrak a_1(t)$ (blue)  and $\mathfrak a_2$ (dashed red) for different values of $\zeta$. As shown here, for $\zeta=0$, $\mathfrak a_1=\mathfrak a_2$ are time independent constants, while $\zeta\neq0$ these quantities are time dependent.}\label{fig:num2}
\end{figure*}

\begin{equation}
\beta=\frac{1}{\sqrt{\hat N^2-\frac{(\hat E^2)^2}{\hat E^1}(\hat N^x)^2}}.
\end{equation}
The mass $m_d$ is related to $m_e$ and $m_i$ via
\begin{equation}
a m_d=m_e-m_i
\end{equation}
where $a$ is a constant defined by
\begin{equation}
\begin{aligned}
a=&\alpha(m_e)+\alpha(m_i),
\end{aligned}
\end{equation}
with 
\begin{equation}
\alpha(m)=\frac{\sqrt{(\widehat {\partial_tE^1})^2\beta^2-8 \sqrt{\hat E^1} m+4 \hat E^1}}{4 \sqrt{\hat E^1}}.
\end{equation}
Motivated by these relations, we evaluate
\begin{equation}
\mathfrak a_1(t)=\alpha(m_e)+\alpha(m_i),
\end{equation}
and 
\begin{equation}
\mathfrak a_2(t)=\frac{-2 \beta \hat E^2   \hat N (m_e-m_i)}{[\partial_xE^1]\sqrt{\hat E^1}}
\end{equation}
from our numerical results for different values of $\zeta$. For $\zeta=0$, which corresponds to classical GR, the Israel junction condition requires that $\mathfrak a_1(t)$ and $\mathfrak a_2(t)$ coincide and remain equal to the same constant $a$, independently of $t$. For nonzero $\zeta$, these quantities should be understood as diagnostics measuring deviations from the classical relation, rather than as junction conditions imposed on the
effective dynamics. As shown in Fig. \ref{fig:num2}, this is indeed satisfied by our numerical results, providing a nontrivial check of our numerical strategy. For $\zeta\neq0$, these classical-Israel diagnostic quantities no longer remain constant, indicating deviations of the Bardeen effective dynamics from the classical GR Israel relation.

\section{Conclusion and outlook}\label{sec:con}
In this work, we investigated effective quantum gravity coupled to a dust shell, which has recently attracted considerable attention due to the discovery of shell-crossing singularities in such models. Although it has been recognized that some previous investigations suffered from an inappropriate choice of gauge (or coordinates), no systematic method for selecting a consistent gauge has been available. 

We developed a strategy to derive the equations that constrain the choice of gauge. Instead of requiring the explicit shell contribution to the Hamiltonian and diffeomorphism constraints, we assumed only that the shell energy-momentum tensor takes the dust form $T_{\mu\nu}\propto u_\mu u_\nu$, and that the shell couples only to the spacetime metric. Under these assumptions, the equations of motion can be written as effective Einstein equations, from which equations constraining the lapse function and shift vector can be obtained. We then applied the framework to a generally covariant effective Hamiltonian model of quantum gravity,  coupled  to dust shell, whose unique vacuum solution is the Bardeen metric. This model contains a quantum parameter $\zeta$, and classical GR is recovered in the limit $\zeta\to0$.
By comparing our numerical results with those obtained from the Israel junction condition, we find good agreement for $\zeta=0$, providing a nontrivial check of the validity of our approach. For $\zeta\neq0$, by contrast, deviations from the classical Israel junction condition arise due to quantum-gravity effects.

Although our approach can be applied more generally, including cases where the explicit form of the shell contribution to the constraints is not known, its formulation in terms of an effective Einstein equation and the freedom to choose the lapse function and shift vector requires the theory to be generally covariant. Therefore, our strategy does not apply to effective quantum black-hole models that fail to preserve general covariance, such as those considered in \cite{Husain:2022gwp}. At the same time, our analysis clarifies the origin of the difficulties in these previous treatments. In particular, gauges such as the PG or Schwarzschild gauge, when imposed on the entire spatial slice, are incompatible with a dust shell, since they require the areal radius to remain smooth across the shell and therefore lead to ill-defined expressions, as discussed under \eqref{eq:nonareal}.

We finally comment on the relation between our construction and recent treatments of thin shells in effective Hamiltonian models, where the junction conditions are obtained by integrating the effective constraints across the shell without requiring a common differentiable coordinate system across it \cite{Fazzini:2025nse,Bobula:2026zlq}. In our model, as in classical GR, junction conditions can be derived from both the constraints and the evolution equations, and their mutual consistency provides an important check that no additional conditions are missed. When the shell dynamics is inferred solely from the integrated constraints, it is therefore not manifest a priori that the resulting junction conditions exhaust those implied by the full dynamics.
Another issue concerns the regularity of the coordinates across the shell. Although the integrated constraints can accommodate nondifferentiable canonical fields, nonsmooth coordinates may themselves introduce additional distributional contributions. Both issues, however, are difficult to address in the models with deformed covariance considered in Refs.~\cite{Fazzini:2025nse,Bobula:2026zlq}, since doing so would require a common smooth coordinate system across the shell, which generally differs from the coordinates in which the deformed-covariance model is originally formulated. The form of the effective theory in such general coordinates is not known. In the present model, by contrast, general covariance allows both issues to be addressed by working in a common smooth coordinate system across the shell, which can be determined together with the junction conditions.

In this work, we have focused on a quantum gravity model coupled to a dust shell. In the future, it would be interesting to apply this approach to other generally covariant effective models, and to extend the analysis to a dust ball coupled to gravity to investigate the dynamics of shocks generated by shell-crossing singularities. This would make it possible to examine whether phenomena such as a spacelike dust-shell trajectory \cite{Sahlmann:2025fde,Liu:2025fil} are genuine physical predictions of these models, or artifacts arising from an inappropriate choice of coordinates.

\begin{acknowledgements}
C.~Z. thanks Zhoujian Cao and Yupeng Zhang for useful discussions on the numerical calculations in this work. We acknowledge useful discussions with Francesco Fazzini on the coupling of dust shells to effective loop quantum black hole models.
This work was supported by the NSFC with Grants No. 12505055, No. 12605088 and  ``the Fundamental Research Funds for the Central Universities''. D.~Q. acknowledges support from the Sichuan Science and Technology Program (Grant No.~2026NSFSC0747).
\end{acknowledgements}

\appendix

\section{Covariant Hamiltonian associated with the Bardeen metric}\label{app:geteffH}
In this section, we will apply this procedure given in \cite{Zhang:2025ccx} to  construct the effective Hamiltonian constraint associated with the Bardeen metric. 

To begin with,  it is convenient to introduce the following phase-space-dependent variables
\begin{equation}
\begin{aligned}
&s_1=E^1,\ s_2=K_2,\ s_3=\frac{K_1}{E^2},\\
&s_4=\frac{\partial_xs_1}{E^2},\ s_5=\frac{\partial_xs_4}{E^2}.
\end{aligned}
\end{equation}
In terms of these variables, the effective Hamiltonian constraint takes the form
\begin{equation}\label{eq:HeffOnMeff}
H_{\rm eff}=-2E^2\Big[\partial_{s_1}M_{\rm eff}+\frac{\partial_{s_2}M_{\rm eff}}{2}s_3+\frac{\partial_{s_4}M_{\rm eff}}{s_4}s_5+\mathcal R\Big],
\end{equation}
where $\mathcal R$ is an arbitrary function of $s_1$ and $M_{\rm eff}$. The mass function $M_{\rm eff}$ depending on $s_1, s_2, s_4$ satisfies 
\begin{equation}\label{eq:solMeff}
(\partial_{s_2}M_{\rm eff})^2=\frac{\mu s_1}{4}\left[(s_4)^2+\mathcal Z\right],
\end{equation}
where $\mu$ and $\mathcal Z$ are arbitrary functions of $s_1$ and $M_{\rm eff}$. Thus, $H_{\rm eff}$ is characterized by the three functions $\mathcal R$, $\mu$, and $\mathcal Z$, which can be determined by requiring a prescribed metric to be a solution of the model, as shown in \cite{Zhang:2025ccx}.

We now specialize to the Bardeen metric given in Eq. \eqref{eq:bardeen-metric}. Requiring this metric to arise as a solution fixes the characteristic functions to
\begin{equation}\label{eq:bardeen-reconstruction-data}
\begin{aligned}
 &\cR=0,\quad \mu=1,\\
 &\cZ(s_1,\Meff)=-4\left[1-\frac{2\Meff s_1}{(s_1+\zeta^2)^{3/2}}\right].
\end{aligned}
\end{equation}
Substituting these functions into Eq.~\eqref{eq:solMeff} and solving for the mass function yields
\begin{equation}
\label{eq:bardeen-mass-function}
 \MB
 =\cA(s_1)\left(1-\frac{(s_4)^2}{4}\right)
 +\cB(s_1)(s_2)^2
\end{equation}
where $\cA$ and $\cB$ are given in Eq. \eqref{eq:AB-definitions}.
Moreover, we have
\begin{equation}
 \lim_{\zeta\to0}\MB
 =\frac{\sqrt{s_1}}{2}
 \left(1+(s_2)^2-\frac{(s_4)^2}{4}\right),
 \label{eq:mass-gr-limit}
\end{equation}
which is the mass function of GR.

For later use, the derivatives of the functions in
Eq.~\eqref{eq:AB-definitions} are
\begin{equation}
\label{eq:AB-derivatives}
\begin{aligned}
 \cA'(s_1)=&
 \frac{\sqrt{s_1+\zeta^2}(s_1-2\zeta^2)}{4(s_1)^2},\\
 \cB'(s_1)=&
 \frac{s_1(s_1+4\zeta^2)}{4(s_1+\zeta^2)^{5/2}},\\
\cA''(s_1)=&
\frac{-(s_1)^2+4s_1\zeta^2+8\zeta^4}
{8(s_1)^3\sqrt{s_1+\zeta^2}},\\
\cB''(s_1)=&
\frac{-(s_1)^2-8s_1\zeta^2+8\zeta^4}
{8(s_1+\zeta^2)^{7/2}}.
\end{aligned}
\end{equation}
The effective Hamiltonian constraint associated with the Bardeen metric is therefore
\begin{equation*}
 H=-2E^2\left[
 \cA'\left(1-\frac{(s_4)^2}{4}\right)
 +\cB'(s_2)^2+\cB s_2s_3-\frac{\cA}{2}s_5
 \right].
\end{equation*}

\section{Some results for classical GR}\label{app:GR}
\begin{figure*}[!ht]
\centering
\begin{subfigure}
\centering
\includegraphics[width=0.3\textwidth]{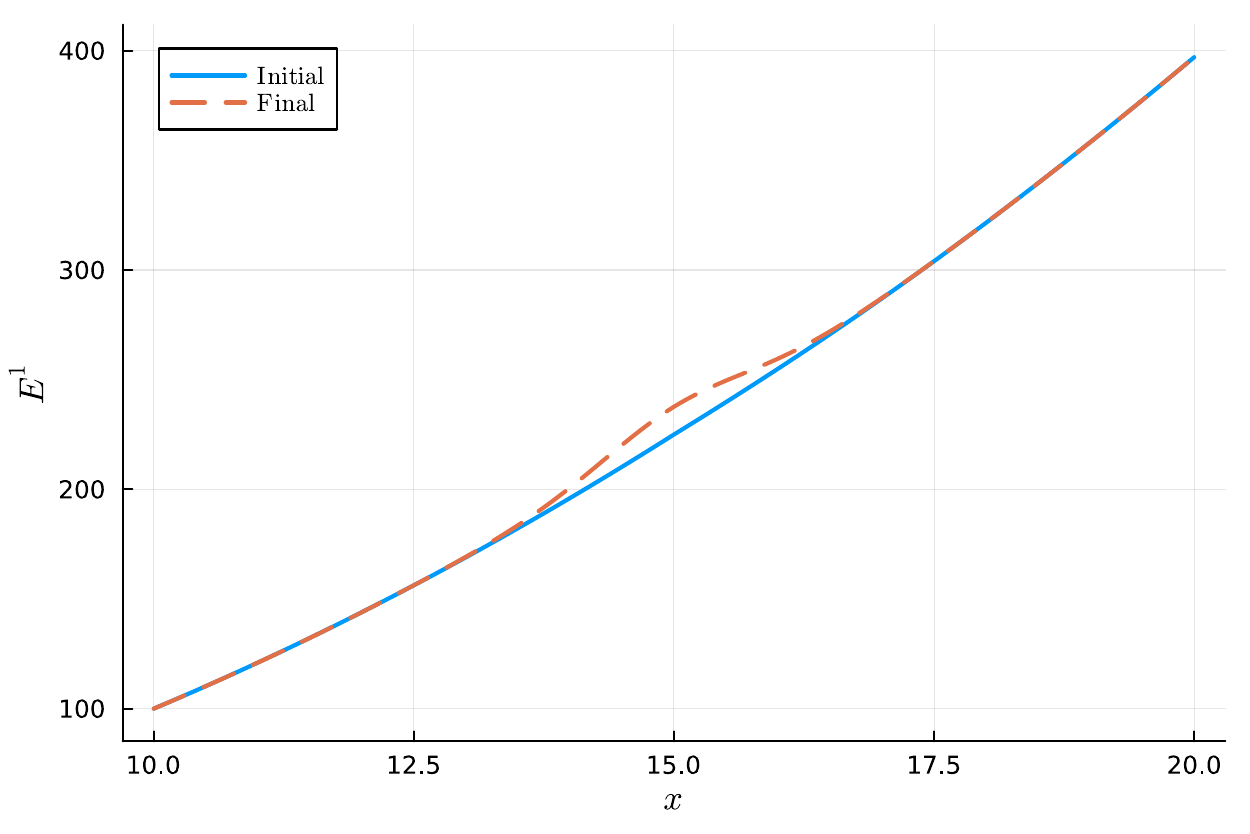}
\end{subfigure}
\begin{subfigure}
\centering
\includegraphics[width=0.3\textwidth]{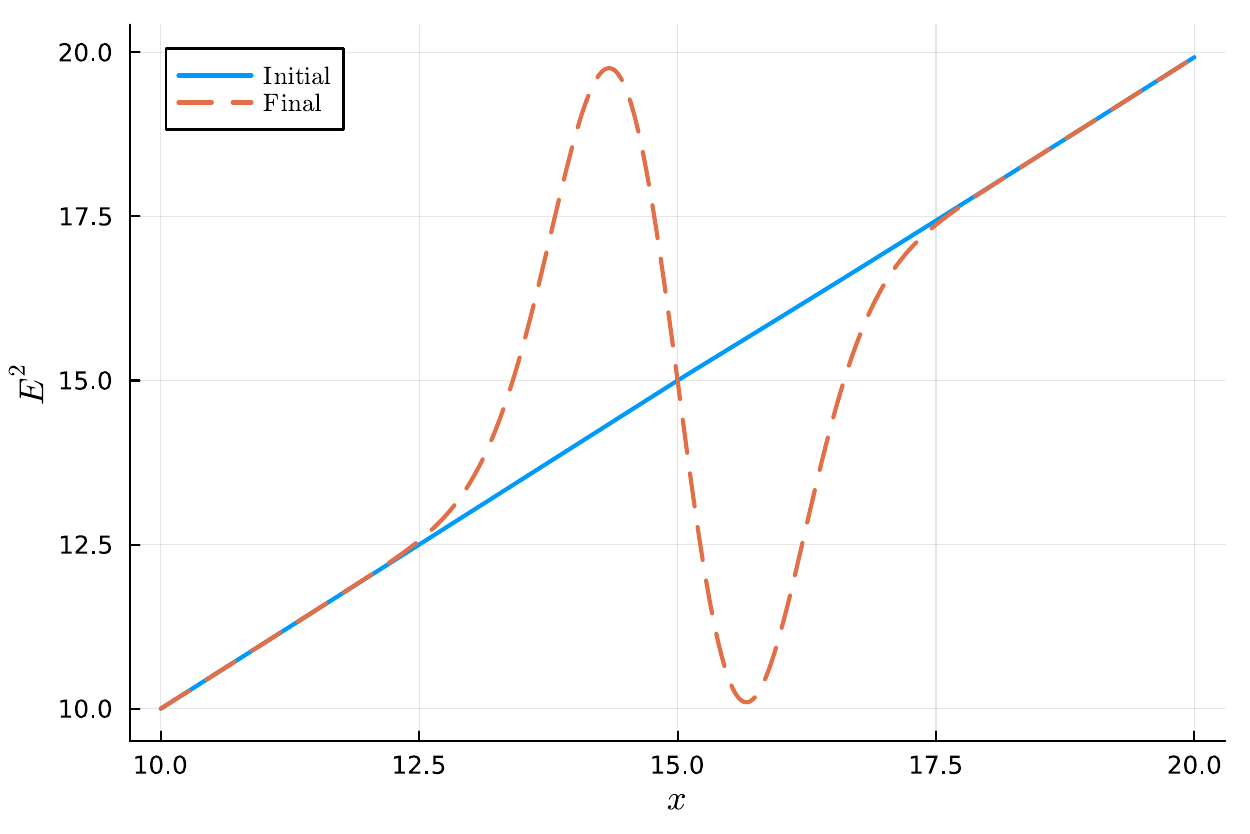}
\end{subfigure}
\begin{subfigure}
\centering
\includegraphics[width=0.3\textwidth]{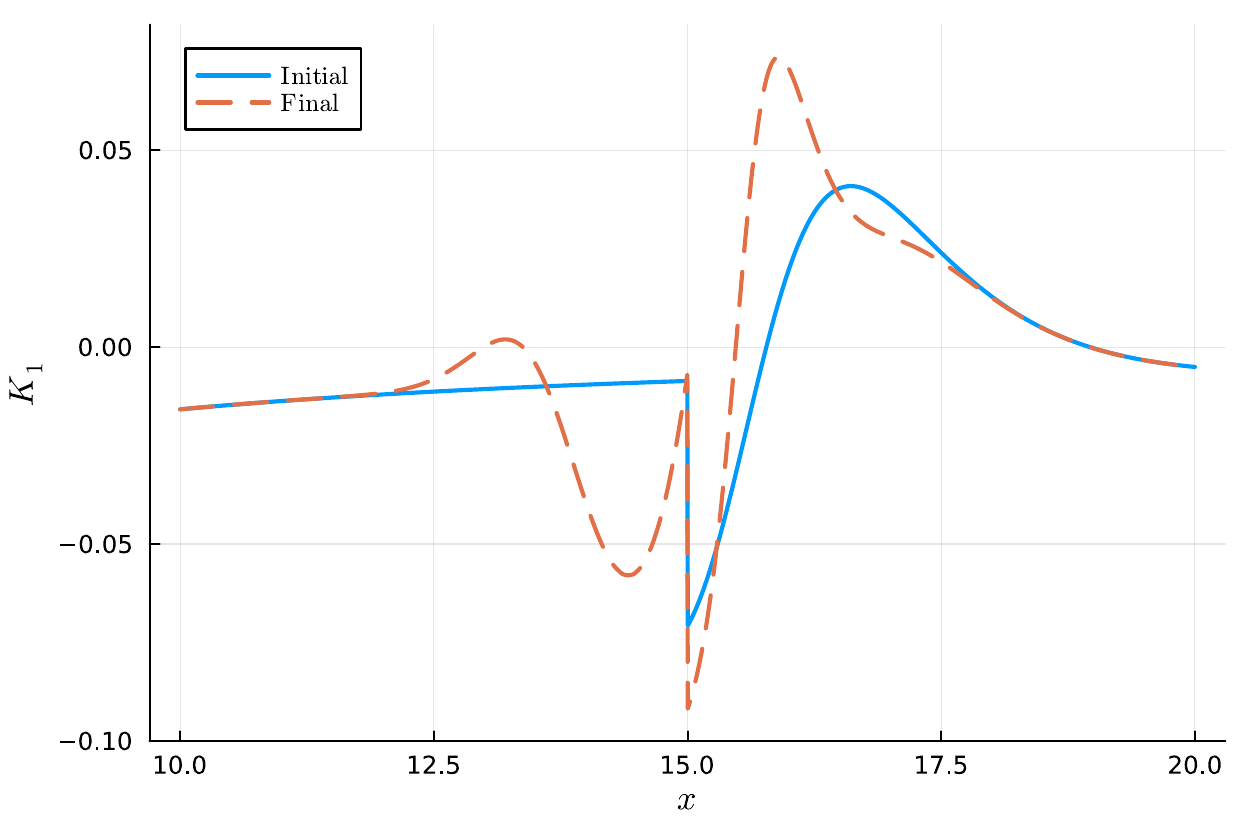}
\end{subfigure}
\begin{subfigure}
\centering
\includegraphics[width=0.3\textwidth]{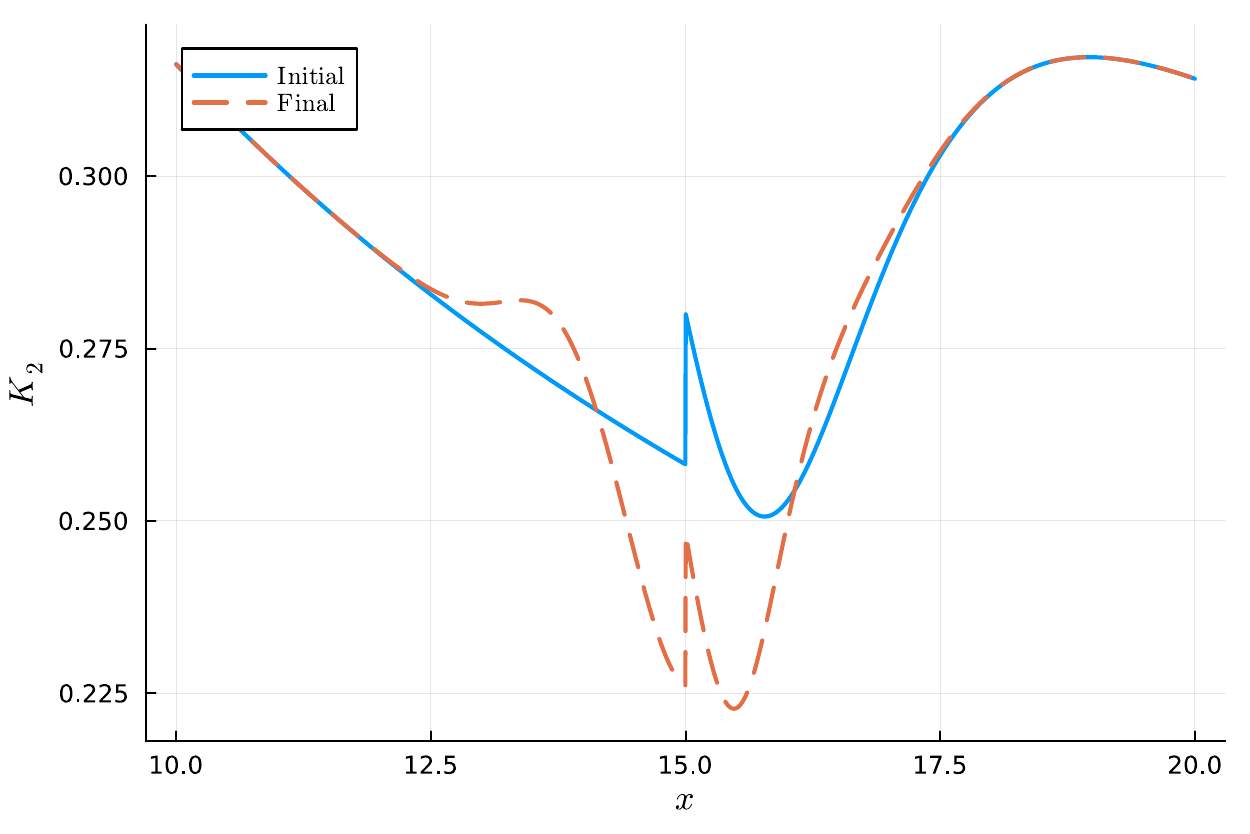}
\end{subfigure}
\begin{subfigure}
\centering
\includegraphics[width=0.3\textwidth]{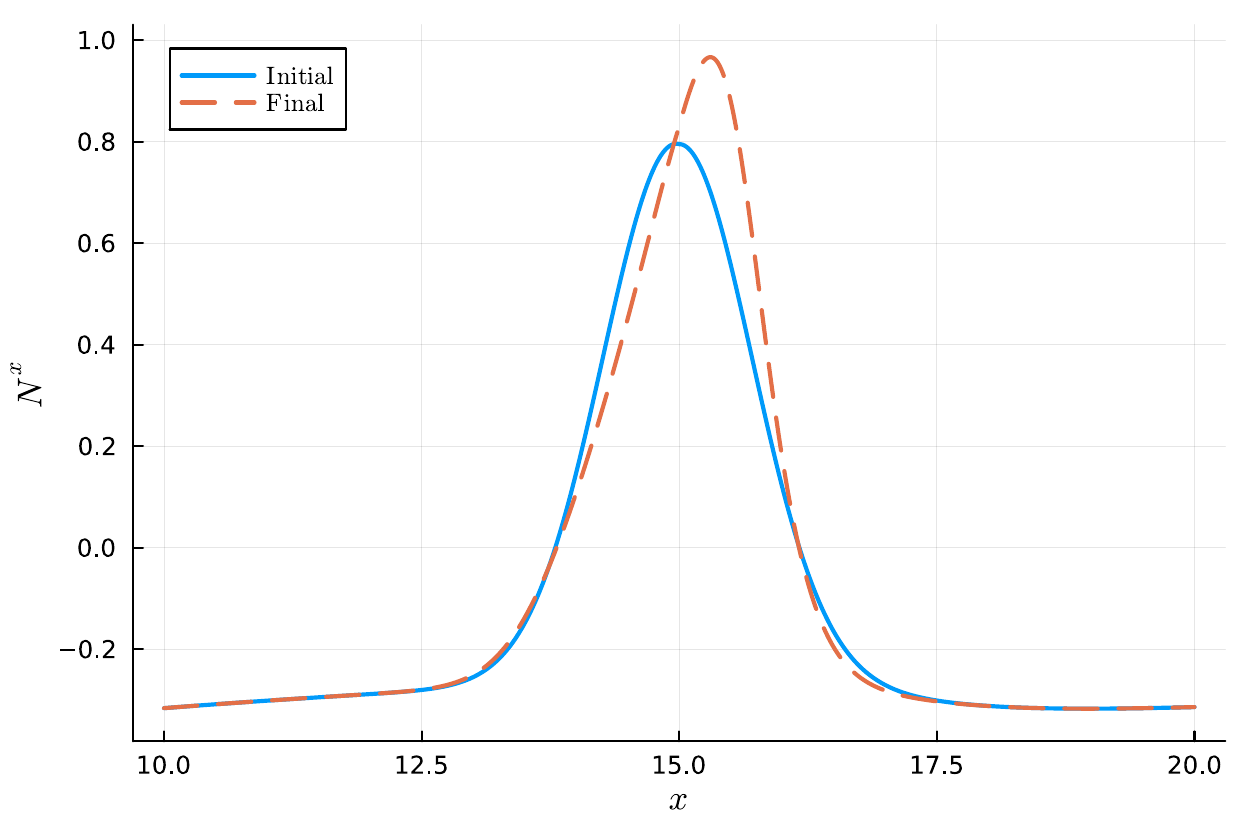}
\end{subfigure}
\begin{subfigure}
\centering
\includegraphics[width=0.3\textwidth]{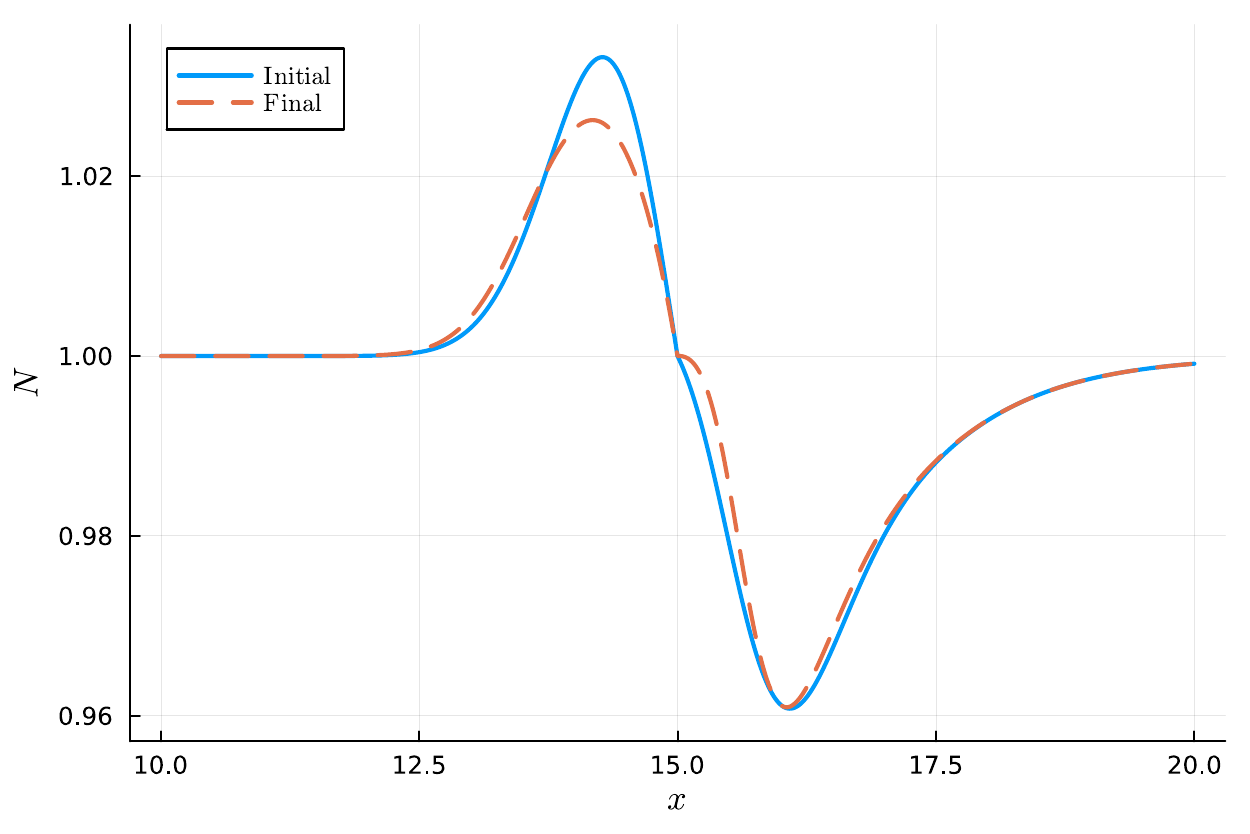}
\end{subfigure}
\caption{Initial (solid blue) and evolved (dashed red) profiles of $E^I$, $K_I$, $N$, and $N^x$ for $m_i=1/2$ and $m_e=1$, shown at $t=0.4$.}\label{fig:num1cl}
\end{figure*}

In classical GR, the Hamiltonian constraint takes the following explicit form:
\begin{equation}
\begin{aligned}
H_{\rm cl}=&\frac{\sqrt{E^1} \partial_x^2E^1}{2 E^2}-\frac{\sqrt{E^1} \partial_xE^1\partial_xE^2}{2 (E^2)^2}+\frac{(\partial_xE^1)^2}{8  \sqrt{E^1}E^2}&\\
&-\frac{E^2(K_2)^2}{2  \sqrt{E^1}}-\frac{E^2}{2  \sqrt{E^1}}-\sqrt{E^1} K_1K_2.
\end{aligned}
\end{equation}
With $H_{\rm cl}$, the Hamiltonian can be written, in analogy with that given in Eq.~\eqref{eq:Hamiton}, as
\begin{equation}
\mathbb H_{\rm cl}=H_{\rm cl}[N]+H_x[N^x].
\end{equation}
Then, the evolution equation for $E^I$ is
\begin{equation}\label{eq:hamiltoneq}
\begin{aligned}
\partial_tE^1&=N^x\partial_xE^1+2 N\sqrt{E^1}K_2,\\
\partial_tE^2&=\frac{N (E^1K_1+E^2K_2)}{\sqrt{E^1}}+\partial_x(N^xE^2),\\
\end{aligned}
\end{equation}
Moreover, the Poisson brackets between $K_I$ and $\mathbb H_{\rm cl}$ read as
\begin{equation}
\begin{aligned}
\{K_1,\mathbb H_{\rm cl}\}=&N\Bigg(-\frac{\partial_xE^1\partial_xE^2}{2 \sqrt{E^1} (E^2)^2}-\frac{(\partial_xE^1)^2}{8\sqrt{E^1}^3E^2}+\\
&\frac{4E^1\partial_x^2E^1}{8\sqrt{E^1}^3E^2}+\frac{E^2 \left(1+(K_2)^2\right)}{2 \sqrt{E^1}^3}-\frac{K_1K_2}{\sqrt{E^1}}\Bigg)+\\
&\frac{\partial_xE^1\partial_xN}{2 \sqrt{E^1} E^2}+\frac{2 E^1\partial_x^2N}{2 \sqrt{E^1} E^2}-\frac{\sqrt{E^1}\partial_xE^2\partial_xN}{(E^2)^2}+\\
&\partial_x(N^xK_1),\\
\{K_2,\mathbb H_{\rm cl}\}=&N\Bigg(\frac{(\partial_xE^1)^2}{8 \sqrt{E^1}(E^2)^2}-\frac{(K_2)^2+1}{2 \sqrt{E^1}}\Bigg)+\\
&\frac{\sqrt{E^1}\partial_xE^1 \partial_xN}{2 (E^2)^2}+N^x\partial_xK_2.
\end{aligned}
\end{equation}
Following the derivation given in Sec.~\ref{sec:jumpcondition}, we obtain the jump condition for classical GR as
\begin{equation}\label{eq:jumpfinalcl}
\begin{aligned}
\hat N^x+\dot\tx=&-\frac{2 \sqrt{\hat E^1} \hat N [K_2]}{[\partial_xE^1]},\\
[\partial_x N^x]=& \frac{2 \sqrt{\hat E^1}\hat N [\partial_xE^2] [K_2]}{\hat E^2 [\partial_xE^1]}-\frac{\sqrt{\hat E^1}\hat N [K_1]}{\hat E^2}-\frac{\hat N [K_2]}{\sqrt{\hat E^1}},\\
[\partial_xN]=& \frac{2 \hat E^2\hat N [K_1] [K_2]}{[\partial_xE^1]}+\frac{2 (\hat E^2)^2\hat N [K_2]^2}{\hat E^1 [\partial_xE^1]}-\frac{\hat N [\partial_xE^1]}{2 \hat E^1},
\end{aligned}
\end{equation}
which coincide with the $\zeta\to0$ limit of the results in Eqs.~\eqref{eq:bardeen-jump-solved-v}–\eqref{eq:bardeen-jump-solved-lapse}.

The numerical evolution of the dust shell coupled to classical GR follows exactly the same procedure as that developed in Sec.~\ref{sec:dyn}. Since the derivation is completely parallel, we do not repeat it here and only list the resulting equations relevant for the numerical calculation:
\begin{itemize}
\item[(1)] In terms of the $s$- and $u$-variables introduced in Sec. \ref{sec:newvar4num}, the evolution equations become:
\begin{equation}\label{eq:gr-conservation}
\partial_t \vec{u} =\partial_x F(\vec{u}) + J(\vec{u})
\end{equation}
with 
\begin{equation}
F=\begin{pmatrix}
s_6\\
\\
s_1s_6\\
\\
s_4s_6+2 N\sqrt{s_1}s_2\\
\\
s_6s_2+\frac{Ns_4\sqrt{s_1}}{2}\\
\\
s_6s_3+s_7
\end{pmatrix},
\end{equation}
and 
\begin{equation}
J=\begin{pmatrix}
\frac{N E^2}{\sqrt{s_1}}\left(s_1s_3+s_2\right)\\
\\
NE^2\sqrt{s_1}(3s_2+s_1s_3)\\
\\
0\\
\\
-\frac{NE^2}{\sqrt{s_1}}\\
\\
\frac{2mN E^2}{ (s_1)^2}
\end{pmatrix}.
\end{equation}
Here we define
\begin{equation}\label{eq:deinem}
m=\frac{\sqrt{s_1}}{2}\left[1+(s_2)^2-\frac{(s_4)^2}{4}\right].
\end{equation} 
\item[(2)]The lapse function and the shift vector are given by
\begin{equation}\label{eq:gr-defineNs6}
\begin{aligned}
N=&g_s\mathring N+\frac{g_b}{2}s_4,\\
s_6=&g_s\mathring s_6-g_b s_2\sqrt{s_1}.
\end{aligned}
\end{equation}
They satisfy  
\begin{equation}\label{eq:jumpfinal01pps}
\begin{aligned}
\frac{\partial_x \mathring s_6}{E^2}=&-\frac{s_2}{\sqrt{s_1}} -\sqrt{s_1}s_3, \\
\frac{\partial_x\mathring N}{E^2}=& \mathring N\left(-\frac{  \mathring s_6}{\sqrt{s_1}^3} \left(s_3s_1+s_2\right) -\frac{ s_4}{2 s_1}\right).
\end{aligned}
\end{equation}
Furthermore, for the scalar $s_7$, we have
\begin{equation}
\begin{aligned}
s_7
=&\sqrt{s_1}\frac{\partial_xg_s}{E^2}\left(\mathring N-\frac{s_4}{2}\right)\\
&+\frac{g_b}{2}\left(\frac{2m}{s_1}+2\sqrt{s_1}s_2s_3\right)+\\
&g_s\mathring N\left(-\frac{  \mathring s_6}{s_1} \left(s_3s_1+s_2\right) -\frac{ s_4}{2 \sqrt{s_1}}\right), 
\end{aligned}
\end{equation}
where we have used $H=0$, which implies
\begin{equation}\label{eq:gr-s5Heffp}
s_5=\frac{2m}{\sqrt{s_1}^3}+2s_2s_3.
\end{equation}
\end{itemize}

The results are shown in Fig. \ref{fig:num1cl}. In this calculation, we choose $g_s(x)=\exp(-(x-\tx_o)^2/\sigma^2)$ with $\sigma=1$, and set the initial radial position of the dust shell to $\tx_o=15$. The initial data are chosen in the same way as described in Sec. \ref{sec:initial}, except that $s_2$ now takes the form
$$
s_2=\pm\sqrt{\frac{2m_o}{\sqrt{s_1}}-1+\frac{(s_4)^2}{4}},
$$
due to the change of the Hamiltonian constraint. The parameters $a$ and $l$ introduced in Eqs.\eqref{eq:ini1} and \eqref{eq:ini2} are chosen to be $l=1.8$ and $a=-0.05$, respectively.

\bibliography{reference}
\end{document}